\documentclass[%
reprint,
amsmath,amssymb,
prb,
]{revtex4-2}

\setcitestyle{numbers,square}
\usepackage{mathrsfs}
\usepackage{comment}

\usepackage{graphicx}% Include figure files
\usepackage{dcolumn}% Align table columns on decimal point
\usepackage{bm}% bold math
\usepackage{blindtext}
\usepackage[normalem]{ulem}

\usepackage{hyperref}
\usepackage{hypcap}
\hypersetup{colorlinks=true,citecolor=blue,linkcolor=blue,urlcolor=blue,filecolor=blue}

\newcommand{\comentar}[1]{}
\newcommand{\vb}[1] {\mathbf{#1}}

\newcommand{\scr}[1] {\mathcal{#1}}
\newcommand{\hb}[1] {\hat{\vb{#1}}}

\newcommand{\ten}[1] {\overleftrightarrow{\vb{#1}}}

\usepackage[dvipsnames]{xcolor} %for more colors

\newcommand{\rme}[0]{{\rm e}}
\allowdisplaybreaks
\begin{document}
%--------------------------------------------%

%%%%%%%%%%%%%%%%%%%%%% TITLE AND ABSTRACT %%%%%%%%%%%%%%%%%%%%%%%%%
%\title{A New Computational Methodology for the Theoretical Study of Linear Momentum Transfer to Nanoparticles Using Electron Beams}
%\title{Linear momentum transfer from a swift electron to a large spherical nanoparticle}
%\title{No net transverse repulsion in causal, multipole-converged momentum transfer from swift electrons to spherical nanoparticles}
%\title{Causal, multipole-converged electrodynamics of swift-electron momentum transfer to spherical nanoparticles: spectral features and no net transverse repulsion}
%\title{Electrodynamics of swift-electron momentum transfer to a large spherical nanoparticle: spectral features and no net transverse repulsion}
\title{Electrodynamics of swift-electron momentum transfer to a large spherical nanoparticle}

%%%%%%%%%%%%%%%%%%%%%%%%%%%%%%%%%%%%%%%%%%%%%%%%%%%%%%%%%%%%%%%%%%%%%%%%%%%
\author{J. Castrej\'on-Figueroa$^{1,2}$}% 
\thanks{These two authors contributed equally}
\email{jcastrejon@ciencias.unam.mx}
\author{J. L. Briseño-Gómez$^1$}
\thanks{These two authors contributed equally}
\email{jorgeluisbrisenio@ciencias.unam.mx}
\author{E. E. Viveros-Armas$^1$}
\author{J. \'A. Castellanos-Reyes$^3$}
\author{A. Reyes-Coronado$^1$}
\affiliation{$^1$Departamento de F\'isica, Facultad de Ciencias, Universidad Nacional Aut\'onoma de M\'exico, Ciudad Universitaria, Av. Universidad $\# 3000$, Mexico City, 04510, Mexico.}
\affiliation{$^2$School of Biotechnology and Biomolecular Science, University of New South Wales, Sydney, NSW, 2052, Australia}
\affiliation{$^3$Department of Physics and Astronomy, Uppsala University, Box 516, 75120 Uppsala, Sweden}
%%%%%%%%%%%%%%%%%%%%%%%%%%%%%%%%%%%%%%%%%%%%%%%%%%%%%%%%%%%%%%%%%%%%%%%%%%%
\date{\today}
%%%%%%%%%%%%%%%%%%%%%%%%%%%%%%%%%%%%%%%%%%%%%%%%%%%%%%%%%%%%%%%%%%%%%%%%%%%
\begin{abstract}
%%%%%%%%%%%%%%%%%%%%%%%%%%%%%%%%%%%%%%%%%%%%%%%%%%%%%%%%%%%%%%%%%%%%%%%%%%%
	Swift electrons from highly focused beams produced in aberration-corrected scanning transmission electron microscopes offer a powerful route for probing and manipulating matter at the nanoscale. Although linear momentum transfer from swift electrons to nanoparticles has been investigated theoretically and experimentally, subsequent analyzes revealed that several earlier predictions relied on non-causal dielectric functions or insufficient numerical convergence, leading to spurious sign reversals in the transferred momentum. 
    Here, we derive analytical expressions and develop a numerically efficient electrodynamic framework to compute the linear momentum transferred from a swift electron to an isolated spherical nanoparticle described by a fully causal, local dielectric response. We apply our framework to large nanoparticles with 50 nm radius and explicitly resolve the spectral density of linear momentum transfer across the full frequency domain. Using causal dielectric functions for aluminum and bismuth, we analyze the role of electron velocity, impact parameter, and material-specific resonances.
   We find that, when causality and full multipolar convergence are enforced, the net transverse linear momentum transferred to spherical nanoparticles remains attractive toward the electron trajectory for all nanoparticles considered, despite the presence of material-dependent sign changes in individual electric and magnetic contributions. These results contrast with earlier theoretical predictions of net repulsive behavior and indicate that additional physical mechanisms beyond the present isolated, local description are required to account for experimentally observed repulsion. Our work establishes a robust reference framework for momentum transfer calculations and provides quantitative benchmarks relevant for electron-beam-based nanoscale manipulation.
\end{abstract}
%%%%%%%%%%%%%%%%%%%%%%%%%%%%%%%%%%%%%%%%%%%%%%%%%%%%%%%%%%%%%%%%%%%%
\maketitle

%%%%%%%%%%%%%%%%%%%%%%%%%%%%%%%%%%%%%%%%%%%%%%%%%%%%%%%%%%%%%%%%%%%%%%%%%%%
\section{Introduction}
%%%%%%%%%%%%%%%%%%%%%%%%%%%%%%%%%%%%%%%%%%%%%%%%%%%%%%%%%%%%%%%%%%%%%%%%%%%

The interaction between fast electrons and matter is at the core of modern electron microscopy and underpins its ability to probe structural, electronic, and optical properties at the nanoscale \cite{Batson0,GarciadeAbajo-1, garcia2021optical, kirkland1998advanced, carter2016transmission}. Advances in aberration correction and electron optics have transformed electron beams into versatile tools capable of delivering spatial resolution approaching the atomic scale and temporal resolution in the femtosecond regime \cite{ Krivanek1, Krivanek2, pennycook2012seeing, dabrowski2020ultrafast, barwick2009, egerton2011electron}. As a result, electron microscopy has evolved from a purely imaging-based technique into a platform for exploring ultrafast electron-matter interactions and nonequilibrium material dynamics \cite{barwick2009, egerton2011electron,zewail2010four}.

Beyond imaging, swift electrons act as broadband, ultrafast sources of electromagnetic fields that can actively drive excitations in nanostructures. Electron energy loss spectroscopy (EELS) provides direct access to these interactions by probing optical, vibrational, and magnetic excitations induced by passing electrons \cite{GarciadeAbajo-1, garcia2021optical}. Recent developments, including four-dimensional EELS (4D-EELS), enable the simultaneous measurement of energy and momentum transfer, offering a more complete characterization of electron-matter coupling \cite{zewail2010four}. When combined with aberration-corrected scanning transmission electron microscopy (STEM), these techniques allow the detection of bright and dark optical modes \cite{Isoniemi2020electron}, phonons and coupled phonon-plasmon excitations \cite{haas2024perspective,mao2025electron, lagos2022advances,gadre2022nanoscale}, and even magnetic excitations such as magnons \cite{kepaptsoglou2025magnon}. In parallel, controlled electron irradiation has been shown to induce atomic rearrangements and directed motion in nanoscale systems \cite{susi2017manipulating, Oleshko, verbeeck2013,Batson01}.

From both experimental and theoretical perspectives, it is now well established that swift electrons are able to transfer linear and angular momentum to nanoparticles (NPs), and that these processes can be described within classical electrodynamics under typical STEM conditions \cite{Batson, GarciadeAbajo0, castrejon2021effects, castellanos2023theory, briseno2024angular}. Early theoretical work demonstrated that aloof electron beams may exert time-averaged forces on small particles comparable in magnitude to those achieved with optical tweezers, suggesting the possibility of electron beam-driven nanoparticle manipulation \cite{GarciadeAbajo0}. Subsequent studies extended this framework to metallic nanoparticles and dimers, revealing a strong dependence of momentum transfer on particle size, material properties, impact parameter, and multipolar excitations \cite{PRBCoronado, Batson, Batson2}. In these works, both attractive and repulsive force components, often associated with the excitation of higher-order multipolar modes, were reported. 

Time-resolved analyzes provided further information by revealing that the interaction between a swift electron and a nanoparticle unfolds on multiple temporal scales \cite{Lagos2, castrejon2021time}. In particular, retardation effects in the induced dipole response can give rise to transient repulsive force components on attosecond timescales, whereas slower electrons experience predominantly attractive interactions. Importantly, these transient force reversals do not necessarily imply a repulsive net momentum transfer when the interaction is integrated over time and frequency. The sign of the total transferred momentum is instead constrained by causality and by the full spectral response of the nanoparticle.

Recent theoretical work has demonstrated that some earlier predictions of net repulsive linear momentum transfer relied on dielectric functions that violate causality or on insufficient numerical convergence in frequency and multipolar expansions \cite{castrejon2021effects}. When causality is enforced and the calculations are fully converged, the net transverse momentum transferred to small spherical nanoparticles was found to be attractive, toward the electron trajectory, in contrast to earlier theoretical claims and with experimental observations of repulsive behavior \cite{Batson}. This discrepancy highlights the need for a robust, causal theoretical framework that serves as a reliable reference and clarifies which physical ingredients are captured within an isolated, local electrodynamic description and which lie beyond it.

In this work, we present a fully causal and numerically converged analysis of the linear momentum transferred from a swift electron to an isolated large spherical nanoparticle of 50~nm radius. Using analytical expressions evaluated for machine precision and causal dielectric functions for aluminum and bismuth \cite{Markovic, werner}, we resolve the spectral density of the transferred linear momentum and quantify the dependence on electron velocity, impact parameter, and material-specific resonances for a nanoparticle with radius $a=50$~nm. Aluminum serves as a canonical Drude-like metal, while bismuth exhibits a complex dielectric response dominated by multiple interband transitions. Our results demonstrate that within a local and causal electrodynamic description, the net transverse linear momentum transferred to spherical nanoparticles remains attractive, despite material-dependent sign changes in individual electric and magnetic contributions. These findings establish a consistent baseline for momentum-transfer calculations and delineate the limits of applicability of isolated-particle models in electron-beam-based manipulation.

%%%%%%%%%%%%%%%%%%%%%%%%%%%%%%%%%%%%%%%%%%%%%%%%%%%%%%%%%%%%%%%%%%%%%%%%%%%%%%%%%
\section{Analytical expressions for the linear momentum transfer}
\label{th}
%%%%%%%%%%%%%%%%%%%%%%%%%%%%%%%%%%%%%%%%%%%%%%%%%%%%%%%%%%%%%%%%%%%%%%%%%%%%%%%%%

In this work, we adopt the local and fully causal dielectric formalism used in Ref.~\cite{castrejon2021effects}, which provides a classical electrodynamic description of the interaction between a swift electron and a spherical nanoparticle, and it is well suited to model linear momentum transfer under typical STEM conditions \cite{castellanos2023theory}. Within this framework, we solve the Maxwell equations, and the material response is described by a causal frequency-dependent dielectric function.

\begin{figure}[h!]
    \centering
    \includegraphics[width=0.5\linewidth]{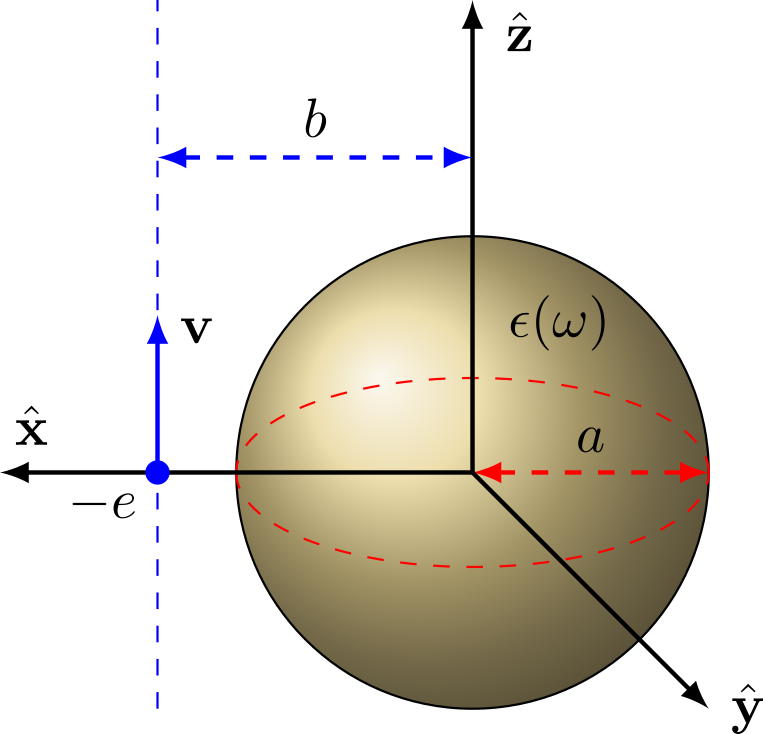}
    \caption{Schematics of the system under study. A spherical nanoparticle of radius $a$, described by a frequency-dependent dielectric function $\epsilon(\omega)$, is embedded in vacuum and centered at the origin. A swift electron (blue dot) travels with constant velocity $\mathbf{v}=v\,\hb{z}$ at an impact parameter $b$ relative to the NP center.} %The planes $x=0$, $y=0$, and $z=0$ are indicated in cyan, yellow, and red, respectively.} 
    \label{fig:system}
\end{figure}

We consider an uncharged nonmagnetic spherical NP of radius $a$, embedded in vacuum and centered at the origin. The NP is characterized by a local dielectric function $\epsilon(\omega)$. A swift electron, modeled as a classical point particle of charge $-e$, travels at constant velocity $\vb{v}=v\,\hat{\vb{z}}$. The impact parameter $b$ is defined as the smallest distance between the electron trajectory and the center of the NP, as illustrated in Fig.~\ref{fig:system}.
The interaction between the electron and the NP is treated using a fully retarded wave solution, in which the total electromagnetic (EM) field is expressed as the superposition of the external field generated by the bare swift electron and the scattered field produced by the electric charge and current densities induced within the NP. Both external and scattered EM fields are expanded in spherical coordinates as multipole series in the frequency domain \cite{de1999relativistic, castrejon2021time}:
\begin{align}
    \vb{E}^{\text{(e,s)}}(\vb{r};\omega)= & \sum_{\ell=1}^{\infty}\sum_{m=-\ell}^{\ell} \Big[
    \scr{E}_{\ell,m}^{\text{(e,s)} r}\,\hb{r} +
    \scr{E}_{\ell,m}^{\text{(e,s)} \theta}\,\hat{\boldsymbol{\theta}} \nonumber \\
    &+
    \scr{E}_{\ell,m}^{\text{(e,s)} \varphi}\,\hat{\boldsymbol{\varphi}}\Big], \label{eq:ScatterdEField} \\
    \vb{H}^{\text{(e,s)}}(\vb{r};\omega)= & \sum_{\ell=1}^{\infty}\sum_{m=-\ell}^{\ell} \Big[ 
    \scr{H}_{\ell,m}^{\text{(e,s)} r}\,\hb{r}+
    \scr{H}_{\ell,m}^{\text{(e,s)} \theta}\,\hat{\boldsymbol{\theta}} \nonumber \\
    &+\scr{H}_{\ell,m}^{\text{(e,s)} \varphi}\,\hat{\boldsymbol{\varphi}}\Big], \label{eq:ScatterdMField} 
\end{align}
where the superscript $\text{(e,s)}$ denotes either the external ($\text{e}$) or scattered ($\text{s}$) field contribution. The explicit expressions for the multipole coefficients $\scr{E}_{\ell,m}^{\text{(e,s)}}$ and $\scr{H}_{\ell,m}^{\text{(e,s)}}$, expressed in SI units, are provided in Appendix~\ref{app: multipole expansion}.

The total linear momentum transferred from the swift electron to the NP is obtained from the conservation of linear momentum. In the frequency domain, the linear momentum transfer along an arbitrary direction $\hb{n}$ can be written as \cite{castrejon2021time}
\begin{equation}
    \Delta{\mathbf P}\cdot\hb{n}=\Delta P_n = \int_0^{\infty} \mathcal{P}_n(\omega)\,d\omega,
\end{equation}
where $\mathcal{P}_n(\omega)$ is the spectral density of the linear momentum transfer.

By symmetry with respect to the $xz$ plane ($y=0$), the $y$ component vanishes. We therefore focus on the transverse component $\Delta P_{\bot}\equiv \Delta P_x$ and the longitudinal component $\Delta P_{\parallel}\equiv \Delta P_z$ defined in Fig.~\ref{fig:system}. In the context of nanoparticle manipulation, the transverse component is of primary interest, as it produces lateral forces acting on the NP. With the geometry of Fig.~\ref{fig:system}, we define $\Delta P_{\bot}>0$ as momentum transfer toward to the electron trajectory (attraction).
%%%%%%%%%%%%%%%%%%%%%%%%%%%%%%%%%%%%%%%%%%%%%%%%%%%%%%%%%%%%%%%%%%%%%%%%%%%%%%%%%
\subsection{The spectral density of the linear momentum}\label{th A}
%%%%%%%%%%%%%%%%%%%%%%%%%%%%%%%%%%%%%%%%%%%%%%%%%%%%%%%%%%%%%%%%%%%%%%%%%%%%%%%%%
The spectral density of the linear momentum transfer is obtained from a closed-surface integral of the Maxwell stress tensor expressed in the frequency domain \cite{castrejon2021time}. We choose the integration surface $S$ as a sphere of radius $R$, centered at the origin, which fully encloses the nanoparticle and does not intersect the electron trajectory. The spectral density of the momentum transfer in the direction $\hb{n}$ is then given by
\begin{equation}
    \mathcal{P}_{n} (\omega) = \frac{R^2}{\pi} \oint_S \hb{n}\cdot \ten{\mathscr{T}} (\vec{r},\omega) \cdot \hb{r} \, d\Omega,
    \label{eq: spectral density LMT}
\end{equation}
where $\ten{\mathscr{T}}(\vb{r},\omega)$ is the Maxwell stress tensor in the frequency domain, defined as \cite{castrejon2021time}
\begin{align} 
& \ten{\mathscr{T}} (\vb{r},\omega) = \, \, \!\Re \bigg\{\!
\epsilon_0 \Big[\vb{E}(\vb{r};\omega)\vb{E}^{*}(\vb{r};\omega) \!-\! \frac{\ten{I}}{2}\vb{E}(\vb{r};\omega) \! \cdot \!\vb{E}^{*}(\vb{r};\omega) \Big] \nonumber \\
& \,\, + \, \mu_0 \Big[\vb{H}(\vb{r};\omega)\vb{H}^{*}(\vb{r};\omega) \! - \! \frac{\ten{I}}{2}\vb{H}(\vb{r};\omega)\! \cdot \!\vb{H}^{*}(\vb{r};\omega) \Big] \! \bigg\}.
\label{eq: maxwell tensor freq}
\end{align}
Here, $\vb{E}$ and $\vb{H}$ denote the total electric and magnetic fields, respectively; $\Re\{\cdot\}$ indicates the real part; $(\cdot)^{*}$ indicates complex conjugation; and $\ten{I}$ is the unit dyadic. The stress tensor naturally separates into electric and magnetic contributions, each of which can be further decomposed into terms arising from the external field, the scattered field, and their interaction. Here, ``electric'' and ``magnetic'' contributions denote the $\epsilon_0[\cdots]$ and $\mu_0[\cdots]$ parts of the Maxwell stress tensor in Eq.~\eqref{eq: maxwell tensor freq}, evaluated from the total fields.

Accordingly, the electric contribution to the stress tensor can be written as
\begin{equation} \label{Tsum}
    \ten{\mathscr{T}}_{\rm E} = \ten{\mathscr{T}}^{\rm ee}_{\rm E}+\ten{\mathscr{T}}^{\rm ss}_{\rm E}+\ten{\mathscr{T}}^{\rm int}_{\rm E},
\end{equation} 
with the interaction term defined as
\begin{equation}\label{Tint}
    \ten{\mathscr{T}}^{\rm int}_{\rm E} = \ten{\mathscr{T}}^{\rm es}_{\rm E}+\ten{\mathscr{T}}^{\rm se}_{\rm E}.
\end{equation}

The interaction term represents the cross contributions between the electron-generated field and the field scattered by the nanoparticle. As will be shown in the following, this interaction term plays a dominant role in the linear momentum transfer and reflects the fundamentally near-field and interference-driven nature of the electron-nanoparticle interaction.

Substituting the multipole expansions of Eq.~\eqref{eq:ScatterdEField} into Eq.~\eqref{Tsum}, the radial component of the electric contribution to the stress tensor can be written as
\begin{align}\label{Tradial}
    \ten{\mathscr{T}}^{(\text{e,s})(\text{e}^{\prime},\text{s}^{\prime})}_{\rm E} &\cdot \hb{r} = \epsilon_0 \sum_{\ell, m}\sum_{\ell^\prime,m^\prime} \Re \Bigg\{ \mathcal{E}_{\ell^\prime, m^\prime}^{(\text{e}^{\prime},\text{s}^{\prime})r*} \Big( \mathcal{E}_{\ell, m}^{(\text{e,s})r} \hb{r} \nonumber \\
    &+ \mathcal{E}_{\ell, m}^{(\text{e,s})\theta} \hat{\boldsymbol{\theta}} + \mathcal{E}_{\ell, m}^{(\text{e,s})\varphi} \hat{\boldsymbol{\varphi}}\Big)  
    - \frac{1}{2}\Big( \mathcal{E}_{\ell, m}^{(\text{e,s})r} \mathcal{E}_{\ell^{\prime}, m^{\prime}}^{(\text{e}^{\prime},\text{s}^{\prime})r*} \nonumber \\
    &+ \mathcal{E}_{\ell, m}^{(\text{e,s})\theta} \mathcal{E}_{\ell^{\prime}, m^{\prime}}^{(\text{e}^{\prime},\text{s}^{\prime})\theta*} + \mathcal{E}_{\ell, m}^{(\text{e,s})\varphi} \mathcal{E}_{\ell^{\prime}, m^{\prime}}^{(\text{e}^{\prime},\text{s}^{\prime})\varphi*} \Big) \hb{r} \Bigg\},
\end{align}
where the superscripts $(\text{e,s})$ and $(\text{e}^{\prime},\text{s}^{\prime})$ again denote external or scattered field contributions.

The products of the electromagnetic field components that appear in Eq.~\eqref{Tradial} can be analytically evaluated. As an example, the product of the radial electric-field components takes the form
\begin{align}
%
% Electric-Electric rr
%
\mathcal{E}_{
\ell, m}^{\text{(e,s)r}}
\mathcal{E}_{\ell^{\prime}, m^{\prime}}^{(\text{e}^{\prime},\text{s}^{\prime})\text{r}\,*}
= \, &D_{\ell, m}^{\rm \text{(e,s)}}D_{\ell^{\prime}, m^{\prime}}^{\rm (\text{e}^{\prime},\text{s}^{\prime}) \, *} \frac{Z_{\ell}^{\rm (e,s)}Z_{\ell^{\prime}}^{\rm (\text{e}^{\prime},\text{s}^{\prime}) \, *}}{(k r)^2} \ell\ell^{\prime}\nonumber\\
&\times (\ell^{\prime}+1)(\ell+1)\nonumber\\
&\times P_{\ell}^m P_{\ell^{\prime}}^{m^{\prime}} \rme^{i (m-m^{\prime}) \varphi},
\label{eq: radial electric-field stress component}
\end{align}
where $Z_\ell^{(\mathrm{e})}=j_\ell(kr)$ ensures regularity at the origin of the external-field expansion, and $Z_\ell^{(\mathrm{s})}=h_\ell^{(+)}(kr)$ enforces the outgoing-wave condition for the scattered field \cite{castrejon2021time}. The coefficients $D_{\ell,m}^{(\text{e,s})}$ are the corresponding multipole amplitudes, defined in the Appendix~\ref{app: multipole expansion}.

Upon substitution into Eq.~\eqref{eq: spectral density LMT}, angular integration over the closed surface can be carried out analytically. For instance, the contribution of the term in Eq.~\eqref{eq: radial electric-field stress component} to the transverse ($\hb{x}$) component of the spectral momentum density is given by
\begin{align}
%
% Electric-Electric rr sin \theta cos \varphi
%
\oint_S \mathcal{E}_{
\ell, m}^{(\text{e,s})\text{r}}
\mathcal{E}_{\ell^{\prime}, m^{\prime}}^{(\text{e}^{\prime},\text{s}^{\prime})\text{r}\,*} \sin\theta\cos\varphi\,d\Omega
= \, &D_{\ell, m}^{\text{(e,s)}}D_{\ell^{\prime}, m^{\prime}}^{(\text{e}^{\prime},\text{s}^{\prime}) \, *} \nonumber\\
&\frac{Z_{\ell}^{(\text{e,s})}Z_{\ell^{\prime}}^{(\text{e}^{\prime},\text{s}^{\prime}) \, *}}{(k r)^2}\nonumber\\
& \times \ell\ell^{\prime}(\ell^{\prime}+1)(\ell+1) \nonumber\\
&\times  \left(\delta_{m+1,m^{\prime}}+\delta_{m-1,m^{\prime}}\right) \nonumber\\
&\times \pi \, IN_{\ell,\ell^{\prime}}^{m,m^{\prime}}.
\end{align}

Here, the quantity
\begin{equation}
    IN_{\ell,\ell^{\prime}}^{m,m^{\prime}} = \int_{-1}^1 P_{\ell}^m (x) P_{\ell^{\prime}}^{m^{\prime}} (x) \sqrt{1-x^2}\,dx,
\end{equation}
is one of the irreducible integrals introduced from now on. These integrals depend only on the integers $\ell,\ell^{\prime},m,m^{\prime}$ of the associated Legendre functions $P_{\ell}^m (x)$ and are independent of the parameters that describe the swift electron or the NP.

Moreover, the irreducible integrals allow results to be free from numerical errors when computed to machine precision using Gaussian quadratures. We refer the reader to Appendix \ref{app: momentum spectral density} for the remaining irreducible integrals that appear in the computation of $\mathcal{P}_{n} $\,\footnote{
The numerical implementation used in this work is openly available on GitHub~\cite{LMTRepo}.
Details on compilation, execution, and data organization are provided in the repository documentation and in Appendix \ref{app:numerical}.}.

%%%%%%%%%%%%%%%%%%%%%%%%%%%%%%%%%%%%%%%%%%%%%%%%%%%%%%%%%%%%%%%%%%%%%%%%%%%%%%%%%
\section{Linear Momentum Transferred to Aluminum and Bismuth Nanoparticles}
%%%%%%%%%%%%%%%%%%%%%%%%%%%%%%%%%%%%%%%%%%%%%%%%%%%%%%%%%%%%%%%%%%%%%%%%%%%%%%%%%

In this Section, we analyze the transverse component of the linear momentum transferred from a swift electron to a spherical nanoparticle (NP), which is the component relevant for electron-beam-based manipulation. We focus on the transverse spectral density $\mathcal{P}_{\bot}(\omega)$ and on the total transverse momentum $\Delta P_{\bot}$, defined with respect to the geometry shown in Fig.~\ref{fig:system}, where the transverse ($\bot$) direction corresponds to the $x$ axis. The results are presented as a function of the electron velocity and the impact parameter for a nanoparticle with radius $a=50$~nm.

Accurate evaluation of momentum transfer requires explicit convergence of the multipole expansion appearing in Eqs.~\eqref{eq:ScatterdEField} and~\eqref{eq:ScatterdMField}. For the nanoparticle size considered here ($a=50$~nm), convergence is achieved by including multipole orders up to $\ell_{\max}=50$, as shown in Appendix~\ref{app: multipole convergence}.  The analytical expressions derived in Section ~\ref{th} allow $\mathcal{P}_{\bot}(\omega)$ to be evaluated to machine precision, and the subsequent frequency integration yields $\Delta P_{\bot}$.

%Throughout this Section, we emphasize the transverse momentum transfer, as the longitudinal component $\Delta P_{\parallel}$ is more directly connected to energy-loss processes probed by EELS. For completeness, representative results for the longitudinal momentum transfer are provided in Appendix~\ref{app: momentum parallel}.

While we emphasize the transverse response because of its direct relevance to manipulation, the longitudinal component $\Delta P_{\parallel}$ is also of independent interest, as it is more directly connected to the energy and momentum exchange processes probed by EELS. Representative results for $\mathcal{P}_{\parallel}$ are provided in Appendix~\ref{app: momentum parallel}.

We consider two materials with markedly different dielectric responses. Aluminum is modeled using a Drude dielectric function and serves as a canonical plasmonic metal. Bismuth is described by a causal dielectric function composed of a Drude term and multiple Lorentz oscillators \cite{werner}, capturing its complex interband structure; the parameters of the Lorentzian fit are reported in the Appendix~\ref{app: dielectric Bi}. This choice allows us to contrast momentum transfer in simple and complex materials while maintaining a fully causal description. Size corrections to the dielectric function are neglected. This approximation is well justified in the present context: previous work \cite{castellanos2023theory} demonstrated that such corrections are negligible---in the linear and angular momentum transfer problem---for nanoparticles with radii of order $a\sim 1$~nm, and their influence is therefore expected to be even less significant for the larger particles ($a=50$~nm) considered here.

Details of the numerical implementation, as well as additional dielectric functions already implemented in the code, are documented in the GitHub repository~\cite{LMTRepo}, with a brief summary provided in Appendix~\ref{app:numerical}.

%%%%%%%%%%%%%%%%%%%%%%%%%%%%%%%%%%%%%%%%%%%%%%%%%%%%%%%%
\subsection{Momentum transferred to a Drude-like aluminum nanoparticle}
%%%%%%%%%%%%%%%%%%%%%%%%%%%%%%%%%%%%%%%%%%%%%%%%%%%%%%%%
We first analyze the transverse linear momentum transferred to an aluminum nanoparticle modeled by a Drude dielectric function with parameters $\hbar\omega_p = 13.14$~eV and $\hbar\Gamma = 0.197$~eV~\cite{Markovic}. Because of its simple free-electron response, aluminum provides a convenient reference system for elucidating the roles of multipolar resonances, electron velocity, and impact parameter in the momentum-transfer process. As shown in Appendix~\ref{app: multipole convergence}, convergence with respect to multipole order is essential for accurately resolving the full spectral structure of momentum transfer.
\begin{figure*}
\centering
\includegraphics[width=.85\textwidth]{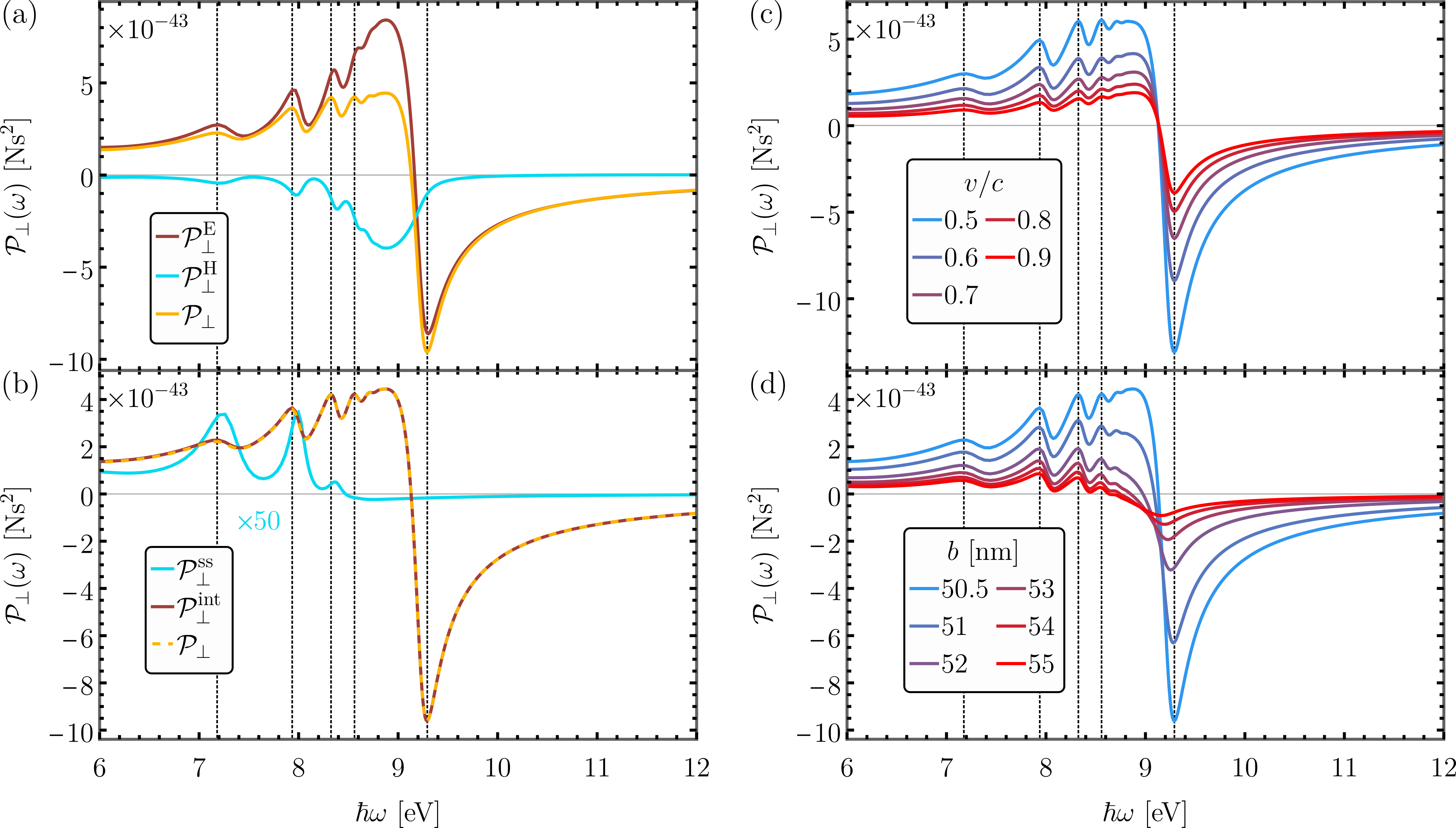}
    \caption{Transverse spectral density of linear momentum $\mathcal{P}_{\bot}(\omega)$ transferred by a swift electron to an aluminum nanoparticle with radius $a=50$~nm. (a) Total spectral density (yellow) decomposed into electric (red) and magnetic (cyan) contributions for $b=50.5$~nm and $v=0.58c$ ($120$~keV). (b) Total spectral density (yellow) decomposed into interaction (external-scattered, red) and scattered-scattered (cyan, scaled by a factor of 50) contributions. The external-external term vanishes identically and it is not shown. (c) $\mathcal{P}_{\bot}(\omega)$ for fixed $b=50.5$~nm and electron velocities indicated in the inset. (d) $\mathcal{P}_{\bot}(\omega)$ for fixed $v=0.58c$ and impact parameters indicated in the inset. Vertical dashed lines indicate representative plasmonic resonances.   
    }
\label{fig: Al_LMED}
\end{figure*}

Figures~\ref{fig: Al_LMED}(a) and~\ref{fig: Al_LMED}(b) show the transverse spectral density of linear momentum, $\mathcal{P}_{\bot}(\omega)$, transferred by a swift electron---with velocity $v=0.58c$ (corresponding to a $120$~keV STEM beam) and impact parameter $b=50.5$~nm---to an aluminum nanoparticle with radius $a=50$~nm. The spectral structure reflects the excitation of different multipolar plasmonic modes by the broadband electromagnetic field associated with the passing electron.

In Fig.~\ref{fig: Al_LMED}(a), the transverse spectral density $\mathcal{P}_{\bot}(\omega)$ is decomposed into its electric $\mathcal{P}^{\rm E}_{\bot}(\omega)$ and magnetic $\mathcal{P}^{\rm H}_{\bot}(\omega)$ contributions. The electric component dominates over most of the frequency range, while the magnetic contribution becomes appreciable near the surface plasmon resonance. For large spherical nanoparticles, the surface plasmon frequency approaches the asymptotic limit $\hbar \omega_s = \hbar\omega_p/\sqrt{2}$~\cite{Bohren}, with retardation effects leading to a modest redshift. In the present geometry, a prominent feature appears near $\hbar\omega \approx 9.3$~eV, close to $\hbar \omega_s$, and it coincides with an enhanced magnetic contribution.

Notably, the resonances observed in the transverse momentum spectral density do not directly correspond to the dominant multipolar amplitudes appearing in Mie scattering or extinction efficiencies (see Appendix~\ref{app: ext and scat mie}). Whereas Mie efficiencies quantify the far-field radiative strength of individual multipoles, the momentum spectral density is governed by the interference between the electron-generated field and the nanoparticle-induced response, making it sensitive to both phase and near-field effects. For comparison, electron energy loss spectroscopy (EELS) and cathodoluminescence (CL) spectra are also presented in the Appendix~\ref{app: ext and scat mie}, where the primary resonance of EELS occurs at the surface plasmon frequency $\omega_s$. As a result, $\mathcal{P}_{\bot}(\omega)$ exhibits additional peaks and enhanced features that are absent in the scattering and extinction spectra. Linear momentum transfer therefore probes near-field and retardation effects that are not captured by conventional far-field observables, providing complementary insight into the electrodynamic response of nanoparticles interacting with swift electrons.

Figure~\ref{fig: Al_LMED}(b) shows $\mathcal{P}_{\bot}(\omega)$ decomposed into contributions arising from the dominant interaction (external-scattered, $\mathcal{P}_{\bot}^{\rm int}$) term and the negligible scattered-scattered ($\mathcal{P}_{\bot}^{\rm ss}$) term of the Maxwell stress tensor, the latter being scaled for visibility. The external-external contribution ($\mathcal{P}_{\bot}^{\rm ee}$) vanishes identically \cite{castellanos2023theory} and is therefore omitted. 

The interaction term of Figure~\ref{fig: Al_LMED}(b) (red curve) accounts for the dominant contribution to the spectral density of transverse momentum $\mathcal{P}_{\bot}$. The total linear momentum associated with the scattered-scattered contribution is negligible. Most of the linear momentum transferred to the nanoparticle therefore originates from the interaction term of the stress tensor, which contains the cross terms between the external electromagnetic field of the swift electron and the electromagnetic fields scattered by the induced electric charges and currents within the nanoparticle. These results demonstrate that the linear momentum transfer from a swift electron to a nanoparticle is governed primarily by interference between the electromagnetic field of the electron and the scattered fields by the nanoparticle, highlighting the fundamental near-field nature of the interaction.

More generally, as shown in the Appendix~\ref{app: ext and scat mie}, the EELS spectrum exhibits a structure similar to that of $\mathcal{P}_{\bot}(\omega)$, while the CL spectrum more closely follows the behavior of $\mathcal{P}_{\bot}^{\mathrm{ss}}(\omega)$.

Figures~\ref{fig: Al_LMED}(c) and~\ref{fig: Al_LMED}(d) examine the dependence of $\mathcal{P}_{\bot}$ on the speed of the electron $v$ and the impact parameter $b$ for an aluminum nanoparticle with radius $a=50$~nm. The spectral-density curves reveal multiple plasmonic resonances whose relative contributions to $\mathcal{P}_{\bot}$ depend sensitively on both $v$ and $b$.

Fig.~\ref{fig: Al_LMED}(c) shows $\mathcal{P}_{\bot}$ for a fixed impact parameter $b=50.5$~nm and electron velocities ranging from $0.5c$ to $0.9c$. Throughout the frequency range, the magnitude of $\mathcal{P}_{\bot}$ decreases with increasing electron velocity, reflecting the reduced interaction time between the swift electron and the nanoparticle. This behavior is consistent with trends reported in earlier studies using different theoretical approaches \cite{castellanos2021angular, castellanos2023theory, castrejon2021effects, castrejon2021time}. 

\begin{figure*}
    \centering
    \includegraphics[width=0.85\textwidth]{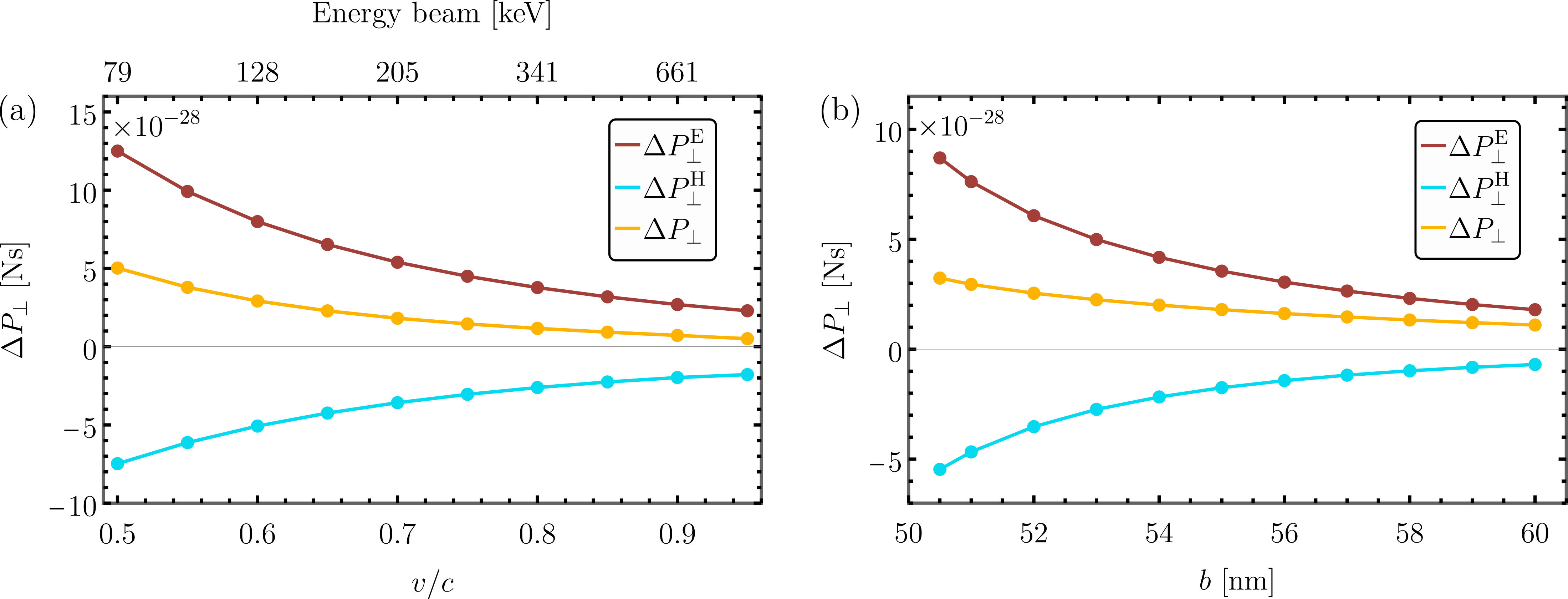}
    \caption{
    Transverse linear momentum ($\Delta P_{\bot}$, yellow dots) transferred by a swift electron to an aluminum NP with $a=50$~nm, (a) as function of $v$ with $b=50.5$~nm, and (b) as function of $b$ with $v=0.58c$ ($120$ keV). The red and blue dots represent the electric and magnetic contributions to $\Delta P_{\bot}$, respectively. The red, blue, and yellow lines are guides to the eye.
    } 
    \label{fig:DPDrudeAlBig}
\end{figure*}

A prior time-resolved analysis~\cite{castrejon2021effects} demonstrated that a small metallic nanoparticle, modeled as an induced electric point dipole, experiences forces on two distinct timescales---attosecond forces of piconewton magnitude and femtosecond forces of attonewton magnitude---during interaction with a swift electron. In that framework, retardation in the induced dipole response gives rise to transient repulsive forces, whereas in the quasistatic limit the interaction remains attractive. The model conserves total linear momentum and predicts no radiation-reaction force, consistent with the negligible scattered-scattered contribution identified in the present work. Although those results were obtained within the dipole approximation, the present fully retarded multipolar analysis shows that the net transverse momentum transfer remains attractive when integrated over frequency.

Figure~\ref{fig: Al_LMED}(d) presents $\mathcal{P}_{\bot}$ for a fixed electron velocity $v=0.58c$ ($120$~keV) and impact parameters ranging from $b=50.5$~nm to $55$~nm.
As $b$ increases, the overall magnitude of $\mathcal{P}_{\bot}$ decreases. At the same time, the spectral profile evolves: higher-order multipolar resonances become increasingly pronounced as the electron beam approaches the nanoparticle surface. This trend is observed from the changing relative weight of the resonances at $\hbar\omega_s=9.3$~eV and $\hbar\omega=7.2$~eV, whose ratio decreases from approximately $4.2$ at $b=50.5$~nm to $1.6$ at $b=55$~nm. These results indicate that the surface plasmon resonance becomes progressively dominant as the electron probe is closer to the nanoparticle surface. In this regime, large nanoparticles exhibit a locally planar response, with the surface plasmon governing the transverse momentum spectral density.

To further explore the effects of the observation on $\mathcal{P}_{\bot}$, Fig.~\ref{fig:DPDrudeAlBig} shows the total transverse linear momentum $\Delta P_{\bot}$ transferred to the aluminum nanoparticle as a function of $v$ [Fig.~\ref{fig:DPDrudeAlBig}(a)] and $b$ [Fig.~\ref{fig:DPDrudeAlBig}(b)]. The electric ($\Delta P_{\bot}^{E}$) and magnetic ($\Delta P_{\bot}^{H}$) contributions are shown separately. 

In both panels of Fig.~\ref{fig:DPDrudeAlBig}, $\Delta P_{\bot}$ decreases with increasing electron velocity and impact parameter. The electric and magnetic contributions act in opposite directions: the electric term is positive, corresponding to attraction of the nanoparticle toward the electron trajectory, while the magnetic term is negative, indicating a repulsive contribution. This repulsion originates from magnetic multipoles induced in the nanoparticle by the time-dependent electromagnetic field of the swift electron. The interaction between these induced magnetic moments and the external field generates a transverse Lorentz force, analogous to the repulsion between magnetic dipoles. Despite this effect, the electric contribution remains dominant over the entire parameter range considered. As a result, the total transverse momentum transfer is always positive (attractive), in agreement with previous findings for small nanoparticles~\cite{castrejon2021effects}.

%%%%%%%%%%%%%%%%%%%%%%%%%%%%%%%%%%%%%%%%%%%%%%%%%%%%%%%%%%%%%%%%%
\subsection{Linear momentum transferred to a bismuth nanoparticle}
%%%%%%%%%%%%%%%%%%%%%%%%%%%%%%%%%%%%%%%%%%%%%%%%%%%%%%%%%%%%%%%%%
Bismuth is modeled using the bulk dielectric function reported by Werner \textit{et al.}~\cite{werner}, expressed as a superposition of one Drude term and nine Lorentz oscillators accounting for interband transitions. The corresponding parameters are provided in Appendix~\ref{app: dielectric Bi}.

\begin{figure*}
    \centering
    \includegraphics[width=0.85\textwidth]{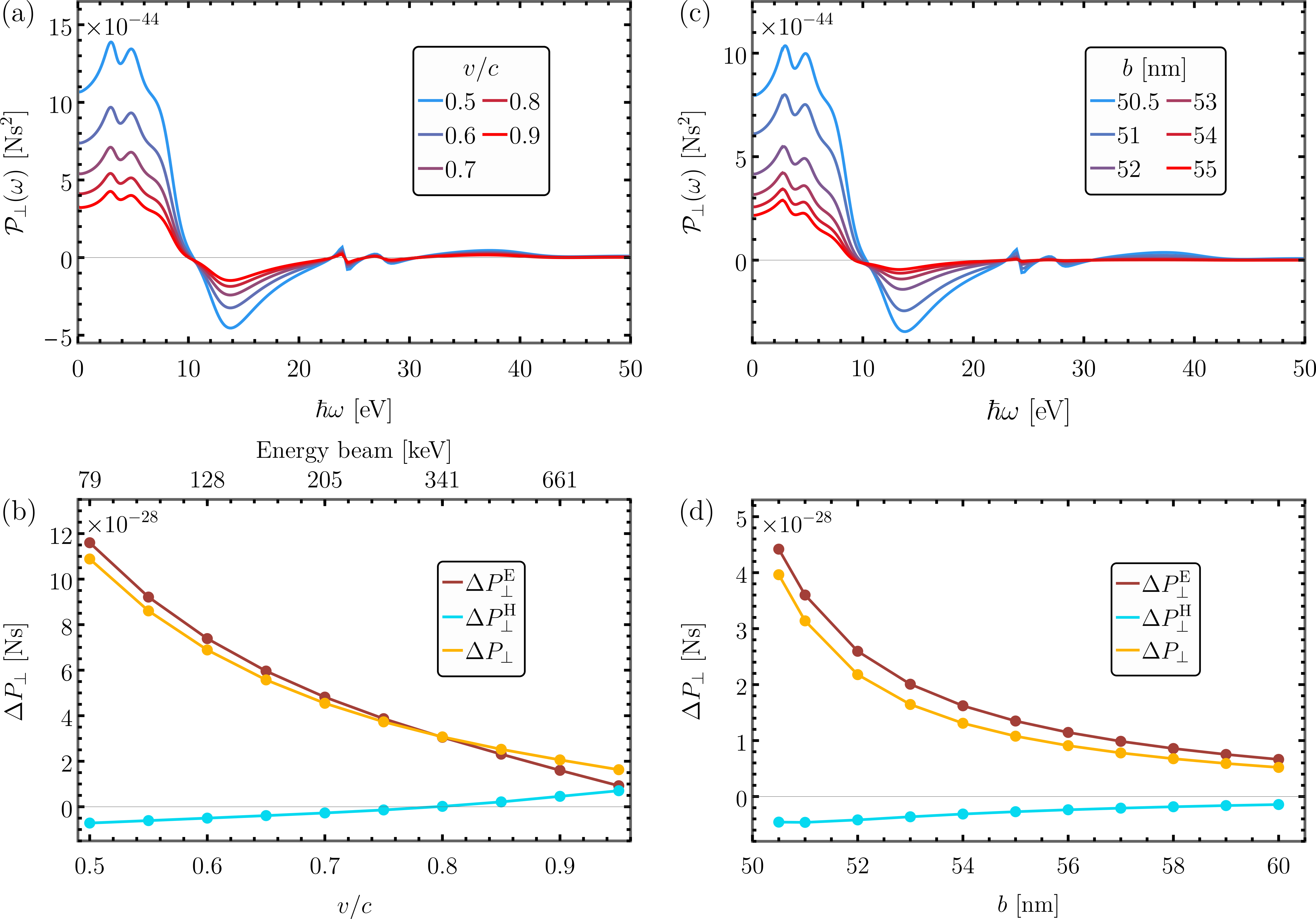}
    \caption{Momentum transfer to a bismuth nanoparticle with radius $a=50$~nm. (a) Transverse spectral density $\mathcal{P}_{\bot}(\omega)$ for fixed impact parameter $b=50.5$~nm and electron velocities indicated in the inset. (b) Total transverse momentum $\Delta P_{\bot}$ as a function of electron velocity for $b=50.5$~nm. (c) $\mathcal{P}_{\bot}(\omega)$ for fixed velocity $v=0.58c$ and impact parameters indicated in the inset. (d) $\Delta P_{\bot}$ as a function of impact parameter for $v=0.58c$. Yellow dots denote total momentum, while red and blue symbols indicate electric and magnetic contributions, respectively. The red, blue, and yellow lines are guides to the eye.}
    \label{fig:Graph vsb Bi}
\end{figure*}

Figure~\ref{fig:Graph vsb Bi}(a) shows the transverse spectral density of linear momentum, $\mathcal{P}_{\bot}(\omega)$, transferred to a bismuth nanoparticle (NP) with radius $a=50$~nm by a swift electron with an impact parameter $b=50.5$~nm, for the electron speeds indicated in the inset. The resonant features observed in $\mathcal{P}_{\bot}(\omega)$ originate from contributions of both free carriers and bound electrons, giving rise to a more intricate spectral structure than that observed for the aluminum NP discussed in the previous section.

As the electron speed increases, the overall magnitude of $\mathcal{P}_{\bot}(\omega)$ decreases due to the reduced interaction time, whereas the relative strength of the resonances remains nearly unchanged. This behavior is similar to the one observed for aluminum nanoparticles. The same tendency is reflected in the total transverse momentum transfer, $\Delta P_{\bot}$, shown as a function of velocity in Fig.~\ref{fig:Graph vsb Bi}(b). The total momentum transfer (yellow dots) remains positive and decreases monotonically with increasing velocity. In contrast, the magnetic contribution (blue symbols) undergoes a sign change, becoming positive at $v \approx 0.78c$, in sharp contrast to the aluminum NP case, where it remains negative over the entire velocity range considered. This sign change highlights that the magnetic contribution is strongly material dependent and sensitive to the dispersive dielectric response. In bismuth, the presence of a pronounced interband structure correlates with a crossover of the magnetic contribution from repulsive to attractive as $v$ increases. By contrast, for Drude aluminum, the magnetic contribution remains negative over the entire velocity range considered.

Figure~\ref{fig:Graph vsb Bi}(c) shows $\mathcal{P}_{\bot}(\omega)$ for a fixed velocity $v=0.58c$ and several impact parameters. For $b=51$~nm, the resonance near $\hbar\omega \approx 3.0$~eV is approximately three times stronger than the resonance at $\hbar\omega \approx 13.7$~eV. As the impact parameter increases, this ratio increases significantly, reaching a value of $6.22$ at $b=55$~nm. This trend does not admit a simple interpretation, since the low-energy resonances do not correspond to dipolar plasmonic modes but rather to interband electronic transitions. Further insight into this behavior is provided in Fig.~\ref{fig:Lorentzian_convergence} of Appendix~\ref{app: multipole convergence}, where the interband transition consistently appears as a resonance at $\omega_0$ across multiple multipole orders.

The response of bismuth NP contrasts with that of aluminum, where the simpler Drude dielectric function allows a more direct interpretation of the dependence of $\mathcal{P}_{\bot}(\omega)$ on the system parameters. Nevertheless, these results highlight that swift electrons probing regions close to the nanoparticle surface carry a rich spectrum capable of efficiently exciting surface and interband modes.

Finally, Fig.~\ref{fig:Graph vsb Bi}(d) shows the total transverse momentum transfer, $\Delta P_{\bot}$, as a function of the impact parameter for the same nanoparticle. As in the aluminum case, $\Delta P_{\bot}$ remains positive and decreases monotonically as $b$ increases. The electric and magnetic contributions act in opposite directions, and the electric contribution dominates in magnitude. Notably, the magnetic contribution is approximately one order of magnitude smaller than the electric one, which may be attributed to the reduced proportion of free carriers in bismuth compared to aluminum.

Overall, the results for aluminum and bismuth demonstrate the versatility of our semi-analytical approach. Simple Drude-like materials allow for a clear identification of multipolar plasmonic resonances and their scaling with electron speed, and impact parameter. In contrast, complex materials such as bismuth exhibit rich spectral features arising from the interplay between plasmonic and interband excitations. Since the present formulation requires only a causal dielectric function as input, it can be readily extended to metals, dielectrics, and semiconductors.

%*********************************************************%
\section{Conclusions}
%*********************************************************%

We have presented a fully causal electrodynamical framework to compute the linear momentum transferred from a swift electron to an isolated spherical nanoparticle. By expressing the spectral density of the momentum transfer in terms of analytical formulations evaluated to machine precision, we have extended previous studies beyond the small-particle limit and provided quantitative results for large nanoparticles of radius of 50~nm. This approach ensures controlled numerical accuracy and allows systematic analysis of the physical mechanisms governing momentum transfer across the entire nanoscale.

Using causal dielectric functions for aluminum and bismuth, we have analyzed the dependence of transverse linear momentum transfer on electron speed, impact parameter, and material-specific resonances. Aluminum, described by a Drude response, exhibits a clear hierarchy of multipolar plasmonic resonances whose relative contributions depend sensitively on the electron trajectory. In contrast, bismuth displays a richer spectral structure arising from the interplay between free-carrier and interband electronic excitations. Despite these differences, a consistent behavior emerges: when causality and full multipolar convergence are enforced, the net transverse linear momentum transferred to spherical nanoparticles remains attractive toward the electron trajectory for all parameters considered.

A key outcome of this work is the explicit identification of the dominant contribution to momentum transfer. We find that the transverse linear momentum arises primarily from the interaction term of the Maxwell stress tensor, which represents near-field interference between the external electromagnetic field of the swift electron and the electromagnetic field scattered by the nanoparticle. In contrast, the contribution associated with scattered fields alone is negligible. This demonstrates that linear momentum transfer in electron–nanoparticle interactions is fundamentally a near-field phenomenon, rather than a radiation-pressure effect governed by far-field scattering or extinction.

Our results also clarify the role of electric and magnetic field contributions. Although the electric component consistently produces an attractive transverse momentum, the magnetic component can be repulsive and may even change sign depending on the material dispersion and electron velocity, as observed for bismuth. Although individual electric or magnetic contributions may exhibit material-dependent sign reversals, their sum yields a net attractive transverse momentum once the full, causally consistent response is taken into account.

Within the assumptions of the present model---namely, isolated spherical
nanoparticles in vacuum described by a local dielectric response---our results demonstrate that net repulsive transverse momentum transfer does not occur. Consequently, experimentally observed repulsive behavior must originate from physical mechanisms not captured by the present model. These may include substrate-mediated forces, nanoparticle charging and secondary-electron emission, nonlocal and quantum-size effects in the dielectric response, thermal gradients, or deviations from spherical geometry.

By establishing a clear and causal baseline for linear momentum transfer, this work delineates the limits of applicability of isolated-particle electrodynamic models and provides quantitative benchmarks for future theoretical and experimental studies. The analytical and computational framework introduced here provides a starting
point for investigating more complex scenarios, including nonspherical, chiral, or magnetic nanoparticles, collective interactions in nanoparticle assemblies, and the simultaneous transfer of linear and angular momentum. Such extensions will be essential for developing predictive models of electron-beam-based nanoscale manipulation and for advancing the concept of electron tweezers beyond idealized systems.
%%%%%%%%%%%%%%%%%%%%%%%%%%%%%%%%%%%%%%%%%%%%%%%%%%%
\begin{acknowledgments}
This work was supported by the UNAM-PAPIIT DGAPA IN101825 project. This research was performed using services/resources provided by Grid UNAM, which is a collaborative effort driven by DGTIC and the research institutes of Astronomy, Nuclear Sciences and Atmosphere Sciences and Climate Change at UNAM, within the framework of the TMERN project. J. C-F. and J. L. B-G. acknowledge PhD scholarships from Secretaría de Ciencia, Humanidades, Tecnología e Innovación (Secihti), Mexico. E. E. V-A acknowledges the master's Scholarship from Secihti, Mexico. J. Á. C-R acknowledges the support of Knut and Alice Wallenbergs' foundation (grant no.\ 2022.0079).
\end{acknowledgments}

%**************************************************
%************* APENDIX ***************************
%************************************************
%%%%%%%%%%%%%%%%%%%%%%%%%%%%%%%%%%%%%%%%%%%
\appendix
%%%%%%%%%%%%%%%%%%%%%%%%%%%%%%%%%%%%%%%%%%%%
\section{Multipole expansion of the external and the scattered electromagnetic fields} \label{app: multipole expansion}

In this Appendix, we provide the explicit expressions for the multipole expansion of the external and scattered electromagnetic fields introduced in Eqs.~\eqref{eq:ScatterdEField} and~\eqref{eq:ScatterdMField}. These expressions constitute the starting point for the analytical evaluation of the Maxwell stress tensor and the subsequent computation of the linear momentum transfer.

Electromagnetic fields are expressed in spherical coordinates $(r,\theta,\varphi)$, related to Cartesian coordinates through $x=r\sin\theta\cos\varphi$, $y=r\sin\theta\sin\varphi$, and $z=r\cos\theta$, with the $x,y$ and $z$ coordinates defined as shown in Fig.~\ref{fig:system} and the nanoparticle located at the origin. 
\begin{align}
    \scr{E}_{\ell,m}^{\text{(e,s)}r}=&\,\rme ^{im\varphi}D_{\ell,m}^{\text{(e,s)}}\ell(\ell+1)P_\ell^m(\cos\theta)\frac{Z_\ell^{\text{(e,s)}}(kr)}{kr},
    \\
%\end{align}
%\begin{align}
    %
    \begin{split}
    \scr{E}_{\ell,m}^{\text{(e,s)}\theta}=&-\rme^{im\varphi}C_{\ell,m}^\text{(e,s)}\frac{m}{\sin\theta}Z_\ell^{\text{(e,s)}}(kr)P_\ell^m(\cos\theta)\\
    &-\rme^{im\varphi}D_{\ell,m}^\text{(e,s)}\left[(\ell+1)\frac{\cos\theta}{\sin\theta}P_\ell^m(\cos\theta)\right.\\
    &-\left.\frac{(\ell-m+1)}{\sin\theta}P_{\ell+1}^m(\cos\theta)\right]\\
    &\times\left[(\ell+1)\frac{Z_\ell^{\text{(e,s)}}(kr)}{kr}-Z_{\ell+1}^{\text{(e,s)}}(kr)\right],
    \end{split}
    \\
%\end{align}
%\begin{align}
    %
    \begin{split}
\scr{E}_{\ell,m}^{\text{(e,s)}\varphi}=&\,i\rme^{im\varphi}C_{\ell,m}^\text{(e,s)}Z_\ell^{\text{(e,s)}}(kr) \\
&\times \left[(\ell+1)\frac{\cos\theta}{\sin\theta}P_\ell^m(\cos\theta)\right.\\
    &\left.-\frac{(\ell-m+1)}{\sin\theta}P_{\ell+1}^m(\cos\theta)\right]\\
    &+i\rme^{im\varphi}D_{\ell,m}^\text{(e,s)}\frac{m}{\sin\theta}P_\ell^m(\cos\theta)\\
    &\times\left[(\ell+1)\frac{Z_\ell^{\text{(e,s)}}(kr)}{kr}-Z_{\ell+1}^{\text{(e,s)}}(kr)\right],
    \end{split}
    %
%\end{align}
%
%and
%
%\begin{align}
   \\ 
   \scr{H}_{\ell,m}^{\text{(e,s)}r}=&\,\rme^{im\varphi}C_{\ell,m}^\text{(e,s)}\ell(\ell+1)P_\ell^m(\cos\theta)\frac{Z_\ell^{\text{(e,s)}}(kr)}{kr},
    \\
    %\end{align}
    %\begin{align}
    %
    \begin{split}
    \scr{H}_{\ell,m}^{\text{(e,s)}\theta}=&\,\rme^{im\varphi}D_{\ell,m}^\text{(e,s)}\frac{m}{\sin\theta}Z_\ell^{\text{(e,s)}}(kr)P_\ell^m(\cos\theta)\\
    &-\rme^{im\varphi}C_{\ell,m}^\text{(e,s)}\left[(\ell+1)\frac{\cos\theta}{\sin\theta}P_\ell^m(\cos\theta)\right.\\
    &-\left.\frac{(\ell-m+1)}{\sin\theta}P_{\ell+1}^m(\cos\theta)\right]\\
    &\times\left[(\ell+1)\frac{Z_\ell^{\text{(e,s)}}(kr)}{kr}-Z_{\ell+1}^{\text{(e,s)}}(kr)\right],
    \end{split}
    \\
    %\end{align}
    %\begin{align}
    %
    \begin{split}
    \scr{H}_{\ell,m}^{\text{(e,s)}\varphi}=&\,i\rme^{im\varphi}C_{\ell,m}^\text{(e,s)}\frac{m}{\sin\theta}P_\ell^m(\cos\theta)\\
    &\times\left[(\ell+1)\frac{Z_\ell^{\text{(e,s)}}(kr)}{kr}-Z_{\ell+1}^{\text{(e,s)}}(kr)\right]\\
    &-\rme^{im\varphi}D_{\ell,m}^\text{(e,s)}Z_{\ell}^{\text{(e,s)}}(kr)  \\
    &\times \left[(\ell+1)\frac{\cos\theta}{\sin\theta}P_\ell^m(\cos\theta)\right.\\
    &-\left.\frac{(\ell-m+1)}{\sin\theta}P_{\ell+1}^m(\cos\theta)\right].
    \end{split}
\end{align}
The superscript $\text{(e,s)}$ denotes either the external or the scattered electromagnetic field, $P_\ell^m$ are the associated Legendre functions, where $Z_\ell^{(\mathrm{e})}=j_\ell(kr)$ ensures regularity at the origin for the external-field expansion, and $Z_\ell^{(\mathrm{s})}=h_\ell^{(+)}(kr)$ enforces the outgoing-wave condition for the scattered field \cite{castrejon2021time}, and $k$ is the vacuum wave number. The coefficients $C_{\ell,m}^{\text{(e,s)}}$ and $D_{\ell,m}^{\text{(e,s)}}$ are defined as
\begin{align}
    C_{\ell,m}^\text{(e,s)}&=(i)^\ell\sqrt{\frac{2\ell+1}{4\pi}\frac{(\ell-m)!}{(\ell+m)!}}\psi_{\ell,m}^\text{M,(e,s)},\\
    D_{\ell,m}^\text{(e,s)}&=(i)^\ell\sqrt{\frac{2\ell+1}{4\pi}\frac{(\ell-m)!}{(\ell+m)!}}\psi_{\ell,m}^\text{E,(e,s)},
\end{align}
where the explicit expressions for the coefficients $\psi_{\ell,m}^{\text{E,e}}$, $\psi_{\ell,m}^{\text{M,e}}$, $\psi_{\ell,m}^{\text{E,s}}$, and $\psi_{\ell,m}^{\text{M,s}}$ can be found in Ref.~\cite{de1999relativistic}.

%%%%%%%%%%%%%%%%%%%%%%%%%%%%%%%%%%%%%%%%%%%%%%%%%%%%%%%%%%%%%%%%%%%%%%%%%%%%%%%%%%%%%%%%%
\section{Expressions for the spectral density of the linear momentum transfer} \label{app: momentum spectral density}

In this Appendix, we provide the explicit expressions required to evaluate the spectral density of the linear momentum transfer defined in Eq.~\eqref{eq: spectral density LMT}. These expressions are obtained from the Maxwell stress tensor in the frequency-domain introduced in Eq.~\eqref{eq: maxwell tensor freq}, evaluated on a closed spherical integration surface that encloses the nanoparticle and does not intersect the electron trajectory.

The electric-field contributions to the radial projection of the Maxwell stress tensor can be written as follows. For the Cartesian $x$ component, one finds the following
\begin{align}
%
%   x component
%
\hb{x}\cdot\ten{\mathscr{T}}^{\rm E}_{\rm \text{(e,s)}(\text{e}^{\prime},\text{s}^{\prime})}\cdot\hb{r} =& \,\epsilon_0 \sum_{\ell, m}\sum_{\ell^\prime,m^\prime} \Re \Bigg\{ \frac{1}{2} \Bigg[\Big( \mathcal{E}_{\ell, m}^{\text{(e,s)}r} \mathcal{E}_{\ell^{\prime}, m^{\prime}}^{(\text{e}^{\prime},\text{s}^{\prime})r*} \nonumber\\
&-\mathcal{E}_{\ell, m}^{\text{(e,s)}\theta} \mathcal{E}_{\ell^{\prime}, m^{\prime}}^{(\text{e}^{\prime},\text{s}^{\prime})\theta*}\nonumber\\
 &- \mathcal{E}_{\ell, m}^{\text{(e,s)}\varphi} \mathcal{E}_{\ell^{\prime}, m^{\prime}}^{(\text{e}^{\prime},\text{s}^{\prime})\varphi*}\Big) \sin\theta \nonumber\\
 &+\mathcal{E}_{\ell, m}^{\text{(e,s)}\theta} \mathcal{E}_{\ell^{\prime}, m^{\prime}}^{(\text{e}^{\prime},\text{s}^{\prime})r*} \cos\theta \Bigg]\cos\varphi\nonumber\\ 
&-\mathcal{E}_{\ell, m}^{\text{(e,s)}\varphi} \mathcal{E}_{\ell^{\prime}, m^{\prime}}^{(\text{e}^{\prime},\text{s}^{\prime})r*}\sin\varphi\Bigg\}.
\label{eq: projection Maxwell Tensor}
\end{align}
The $y$ and $z$ projections  can be found analogously. The magnetic-field contributions are obtained by replacing
$\epsilon_0 \rightarrow \mu_0$ and
$\mathcal{E}_{\ell m} \rightarrow \mathcal{H}_{\ell m}$ in Eq. \eqref{eq: projection Maxwell Tensor}. The full expressions for the external electromagnetic fields are given in Eqs.~\eqref{eq:ScatterdEField} and~\eqref{eq:ScatterdMField}.

Upon substitution into Eq.~\eqref{eq: spectral density LMT}, the closed surface integral over the solid angle can be evaluated analytically. The resulting expressions can be written as sums of angular integrals involving products of multipolar field components. Representative terms are listed below.
\begin{widetext}
\begin{align}
%
% Electric-Electric rr sin \theta sin \varphi
%
\oint_S\mathcal{E}_{
\ell, m}^{\text{(e,s)}r}
\mathcal{E}_{\ell^{\prime}, m^{\prime}}^{(\text{e}^{\prime},\text{s}^{\prime})r\,*} \sin\theta\sin\varphi\,d\Omega
= \, &D_{\ell, m}^{\rm \text{(e,s)}}D_{\ell^{\prime}, m^{\prime}}^{\rm (\text{e}^{\prime},\text{s}^{\prime}) \, *} \frac{Z_{\ell}^{\rm \text{(e,s)}}Z_{\ell^{\prime}}^{\rm (\text{e}^{\prime},\text{s}^{\prime}) \, *}}{(k r)^2} \ell\ell^{\prime}(\ell^{\prime}+1)(\ell+1)   \left(\delta_{m+1,m^{\prime}}-\delta_{m-1,m^{\prime}}\right)  i \pi \, IN_{\ell,\ell^{\prime}}^{m,m^{\prime}}
\end{align}
\begin{align}
%
% Electric-Electric rr cos \theta 
%
\oint_S\mathcal{E}_{
\ell, m}^{\text{(e,s)}r}
\mathcal{E}_{\ell^{\prime}, m^{\prime}}^{(\text{e}^{\prime},\text{s}^{\prime})r\,*} \cos\theta\,d\Omega
= \, &D_{\ell, m}^{\rm \text{(e,s)}}D_{\ell^{\prime}, m^{\prime}}^{\rm (\text{e}^{\prime},\text{s}^{\prime}) \, *} \frac{Z_{\ell}^{\rm \text{(e,s)}}Z_{\ell^{\prime}}^{\rm (\text{e}^{\prime},\text{s}^{\prime}) \, *}}{(k r)^2} \ell\ell^{\prime}(\ell^{\prime}+1)(\ell+1)  2 \, \delta_{m,m^{\prime}}  \, IM_{\ell,\ell^{\prime}}^{m,m^{\prime}}
\end{align}
\begin{align}
%
% Electric-Electric r\theta cos\theta cos\varphi
%
\oint_S\mathcal{E}_{
\ell, m}^{\text{(e,s)}\theta}
&\mathcal{E}_{\ell^{\prime}, m^{\prime}}^{(\text{e}^{\prime},\text{s}^{\prime})r\,*}\cos\theta\cos\varphi\,d\Omega
= \, -\pi\,\ell^{\prime}(\ell^{\prime}+1) \left(\delta_{m+1,m^{\prime}}+\delta_{m-1,m^{\prime}}\right)\Bigg\{m \,C_{\ell,m}^{\rm \text{(e,s)}} \, Z_{\ell}^{\rm \text{(e,s)}} \,IW_{\ell,\ell^{\prime}}^{m,m^{\prime}}\nonumber\\
&+D_{\ell, m}^{\rm \text{(e,s)}} \left[(\ell+1)\frac{Z_\ell^{\text{(e,s)}}(kr)}{kr}-Z_{\ell+1}^{\text{(e,s)}}(kr)\right] \bigg[(\ell+1) IV_{\ell,\ell^{\prime}}^{m,m^{\prime}}-(l-m+1)IW_{\ell+1,\ell^{\prime}}^{m,m^{\prime}}\bigg]\Bigg\} \frac{Z_{\ell^{\prime}}^{\rm (\text{e}^{\prime},\text{s}^{\prime})\, *}}{k r} D_{\ell^{\prime}, m^{\prime}}^{\rm (\text{e}^{\prime},\text{s}^{\prime}) \, *},
\end{align}
\begin{align}
%
% Electric-Electric r\theta cos\theta sin\varphi
%
\oint_S\mathcal{E}_{
\ell, m}^{\text{(e,s)}\theta}
&\mathcal{E}_{\ell^{\prime}, m^{\prime}}^{(\text{e}^{\prime},\text{s}^{\prime})r\,*}\cos\theta\sin\varphi\,d\Omega
= \, -i\pi\,\ell^{\prime}(\ell^{\prime}+1) \left(\delta_{m+1,m^{\prime}}-\delta_{m-1,m^{\prime}}\right)\Bigg\{m \,C_{\ell,m}^{\rm \text{(e,s)}} \, Z_{\ell}^{\rm \text{(e,s)}} \,IW_{\ell,\ell^{\prime}}^{m,m^{\prime}}\nonumber\\
&+D_{\ell, m}^{\rm \text{(e,s)}} \left[(\ell+1)\frac{Z_\ell^{\text{(e,s)}}(kr)}{kr}-Z_{\ell+1}^{\text{(e,s)}}(kr)\right] \bigg[(\ell+1) IV_{\ell,\ell^{\prime}}^{m,m^{\prime}}-(l-m+1)IW_{\ell+1,\ell^{\prime}}^{m,m^{\prime}}\bigg]\Bigg\} \frac{Z_{\ell^{\prime}}^{\rm (\text{e}^{\prime},\text{s}^{\prime})\, *}}{k r} D_{\ell^{\prime}, m^{\prime}}^{\rm (\text{e}^{\prime},\text{s}^{\prime}) \, *}.
\end{align}
\end{widetext}
All remaining angular integrals can be derived by following the same procedure, and can be found in Ref. \cite{castrejon2021phdthesis}. Together, these expressions provide a complete analytical evaluation of the linear-momentum-transfer spectral density.

The expressions presented here are included to illustrate the underlying methodology and to identify the set of irreducible integrals that appear in the full expression of the transverse spectral density $\mathcal{P}_\bot$, which points along the $x$ direction in Fig.~\ref{fig:system}. The complete set of these irreducible integrals is:
\vspace{-0.07cm}
\begin{align}
    IN_{\ell,\ell^{\prime}}^{m,m^{\prime}} =& \int_{-1}^1 P_{\ell}^m (x) P_{\ell^{\prime}}^{m^{\prime}} (x) \sqrt{1-x^2}\,dx, 
    \\
    %\end{align}
    %\begin{align}
    IM_{\ell,\ell^{\prime}}^{m,m^{\prime}} =& \int_{-1}^1 P_{\ell}^m (x) P_{\ell^{\prime}}^{m^{\prime}} (x) \,x\,dx, 
    \\
    %\end{align}
    %\begin{align}
    IU_{\ell,\ell^{\prime}}^{m,m^{\prime}} =& \int_{-1}^1 P_{\ell}^m (x) P_{\ell^{\prime}}^{m^{\prime}} (x) \frac{1}{\sqrt{1-x^2}}\,dx, 
    \\
    %\end{align}    
    %\begin{align}
    IV_{\ell,\ell^{\prime}}^{m,m^{\prime}} =& \int_{-1}^1 P_{\ell}^m (x) P_{\ell^{\prime}}^{m^{\prime}} (x) \frac{x^2}{\sqrt{1-x^2}}\,dx, 
    \\
    %\end{align}
    %\begin{align}
    IW_{\ell,\ell^{\prime}}^{m,m^{\prime}} =& \int_{-1}^1 P_{\ell}^m (x) P_{\ell^{\prime}}^{m^{\prime}} (x) \frac{x}{\sqrt{1-x^2}}\,dx, 
    \\
    %\end{align}
    %\begin{align}
    IX_{\ell,\ell^{\prime}}^{m,m^{\prime}} =& \int_{-1}^1 P_{\ell}^m (x) P_{\ell^{\prime}}^{m^{\prime}} (x) \frac{x^2}{1-x^2}\,dx, 
    \\
    %\end{align}
    %\begin{align}
    IY_{\ell,\ell^{\prime}}^{m,m^{\prime}} =& \int_{-1}^1 P_{\ell}^m (x) P_{\ell^{\prime}}^{m^{\prime}} (x) \frac{x}{1-x^2}\,dx, \\
    IZ_{\ell,\ell^{\prime}}^{m,m^{\prime}} =& \int_{-1}^1 P_{\ell}^m (x) P_{\ell^{\prime}}^{m^{\prime}} (x) \frac{x^3}{1-x^2}\,dx, 
    \\
    %\end{align}
    %\begin{align}
    \Delta_{\ell,\ell^{\prime}} =& \int_{-1}^1 P_{\ell}^m (x) P_{\ell^{\prime}}(x) \, dx \nonumber\\ &= \frac{2(\ell+m)!}{(2\ell+1)(\ell-m)!} \delta_{\ell,\ell^{\prime}},
\end{align}
which can be evaluated to machine precision using Gauss-Legendre and Gauss-Chebyshev quadratures \cite{kahaner1989numerical,press2007numerical}.

The integral of $\mathcal{P}_\bot$ on the frequency space is performed employing the Gauss-Kronrod quadrature, as described in Ref.~\cite{castrejon2021effects}.

%%%%%%%%%%%%%%%%%%%%%%%%%%%%%%%%%%%%%%%%%%%%%%%%%%%%%%%%%%%%%%%%%%%
\section{Multipole Convergence}\label{app: multipole convergence}
%%%%%%%%%%%%%%%%%%%%%%%%%%%%%%%%%%%%%%%%%%%%%%%%%%%%%%%%%%%%%%%%%%%%

All results reported in the main text are obtained from the fully retarded multipole expansions of the external and scattered fields in Eqs.~\eqref{eq:ScatterdEField} and~\eqref{eq:ScatterdMField}, together with the Maxwell stress tensor expression for the spectral density in Eq.~\eqref{eq: spectral density LMT}. In practice, these formulas involve (i) infinite sums over multipole order \(\ell\) and azimuthal index \(m\) in the field expansions, and (ii) corresponding double sums over \((\ell,m)\) and \((\ell',m')\) when forming the quadratic field products that appear in the stress tensor, e.g. Eq.~\eqref{Tradial} and Eq.~\eqref{eq: projection Maxwell Tensor}. Numerical evaluation therefore requires truncating the multipole sums at a maximum order \(\ell_{\max}\), i.e.
\(\sum_{\ell=1}^{\infty}\rightarrow\sum_{\ell=1}^{\ell_{\max}}\)
(and analogously for \(\ell'\)).
The purpose of this Appendix is to document the multipole convergence of the transverse spectral momentum-transfer density \(\mathcal{P}_{\bot}(\omega)\) with respect to this truncation.

Throughout this convergence test, we use the same geometry as in the main text: a spherical nanoparticle of radius \(a=50~\mathrm{nm}\) in vacuum---see Fig.~\ref{fig:system}---, an integration sphere \(S\) that encloses the particle and does not intersect the electron trajectory, and a representative electron trajectory with \(b=50.5~\mathrm{nm}\) and \(v=0.7c\) (the same parameters used in Fig.~\ref{fig: Al_LMED}). We evaluated \(\mathcal{P}_{\bot}(\omega)\) for increasing values of \(\ell_{\max}\) and monitored changes in both the amplitude and the resonant structure of the spectrum.

%\subsection{Drude aluminum}

For aluminum we use the Drude dielectric function with parameters
\(\hbar\omega_p=13.14~\mathrm{eV}\) and \(\hbar\Gamma=0.197~\mathrm{eV}\) taken from Markovi\'c and Raki\'c~\cite{Markovic}, as in the main text.
In this case \(\mathcal{P}_{\bot}(\omega)\) exhibits multiple features across a broad frequency range, reflecting the excitation of several multipolar modes by the electron field.
Here, ``higher multipole orders'' refer to the need to include multipoles beyond the lowest orders (dipole \(\ell=1\), quadrupole \(\ell=2\), etc.).
As \(\ell_{\max}\) increases, additional peaks emerge, and existing peaks change until the spectrum stabilizes.

\begin{figure}[h]
    \centering
    \includegraphics[width=0.9\linewidth]{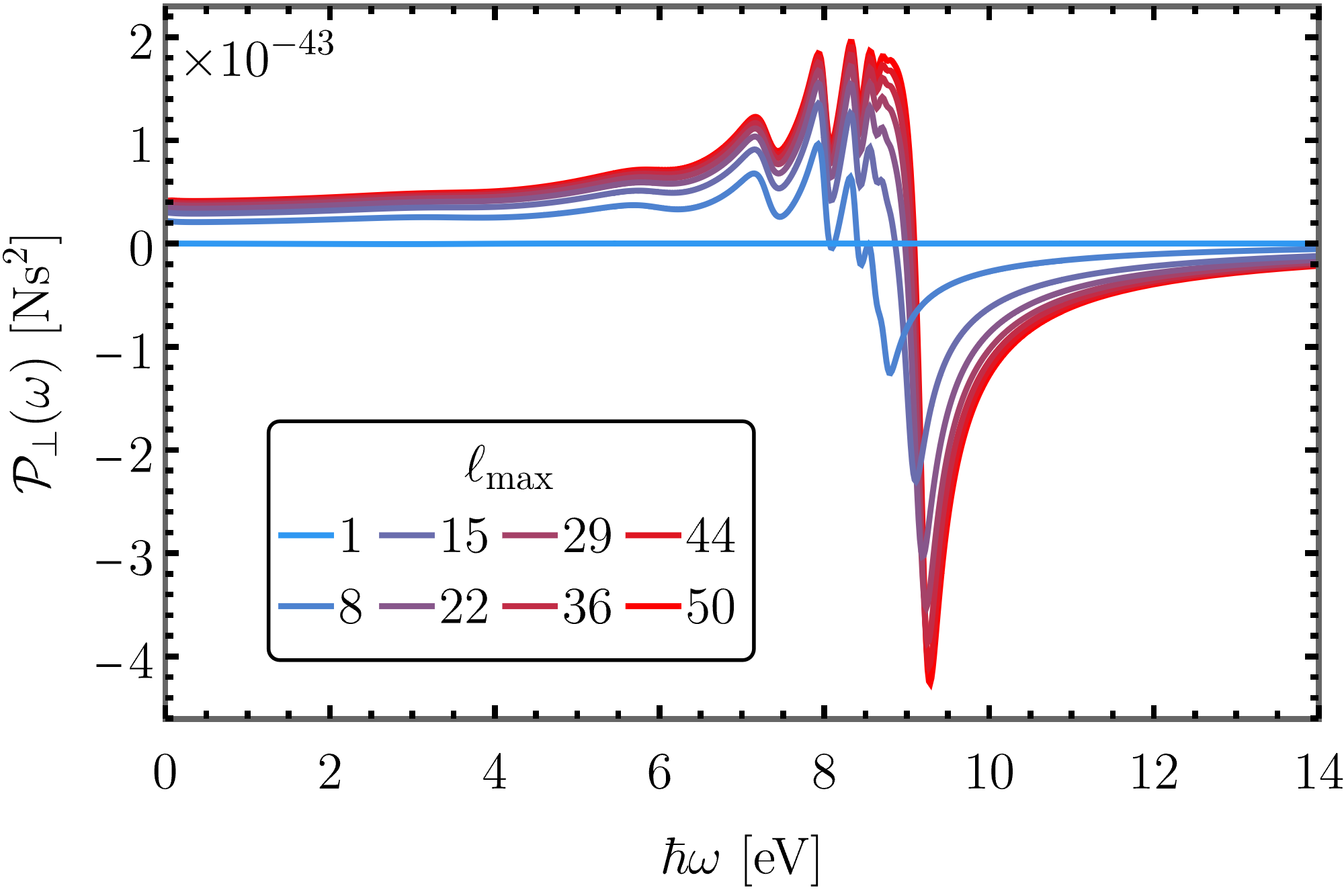}
    \caption{Convergence of the transverse spectral momentum-transfer density \(\mathcal{P}_{\bot}(\omega)\) with respect to the multipole truncation \(\ell_{\max}\) for a Drude aluminum nanoparticle with \(a=50~\mathrm{nm}\), \(b=50.5~\mathrm{nm}\), and \(v=0.7c\). As \(\ell_{\max}\) increases, additional multipolar contributions modify the spectrum until convergence is reached. The spectrum is converged for \(\ell_{\max}=47\); larger values do not produce visible changes.}
    \label{fig:Drude_convergence}
\end{figure}

Figure~\ref{fig:Drude_convergence} provides a visual guide to the convergence of $\mathcal{P}_{\bot}(\omega)$ as progressively higher multipole orders are included up to $\ell_{\max}$. The spectrum is converged for \(\ell_{\max}=47\); higher values do not produce visible changes.

%\subsection{Single-Lorentzian test material}

Since realistic dielectric functions can contain both Drude and Lorentz
contributions---as in the case of bismuth, considered in the main text---it is useful
to analyze the convergence properties of an isolated Lorentz resonance. To this
end, we consider a causal dielectric function consisting of a single Lorentz oscillator:
\begin{equation}
\epsilon(\omega)
=
1+\frac{A}{\omega_0^2-\omega^2-i\omega\Gamma},
\end{equation}
with the resonance frequency fixed at $\hbar\omega_0=24.1~\mathrm{eV}$. This value is
taken directly from the bismuth parametrization listed in
Table~\ref{tab:bismuth_params} (oscillator $n=5$) and is used here only as a
representative resonance frequency.

This simplified model is not intended to reproduce the full dielectric response of bismuth. Rather, by fixing the resonance frequency by construction, it provides a controlled setting in which the dependence on $\ell_{\max}$ can be attributed exclusively to the progressive inclusion of higher-order multipolar contributions. Variations with $\ell_{\max}$ therefore modify the distribution of spectral weight among multipoles and the detailed line shape (peak amplitude and broadening), while leaving the resonance position unchanged.

\begin{figure}[h]
    \centering
    \includegraphics[width=0.9\linewidth]{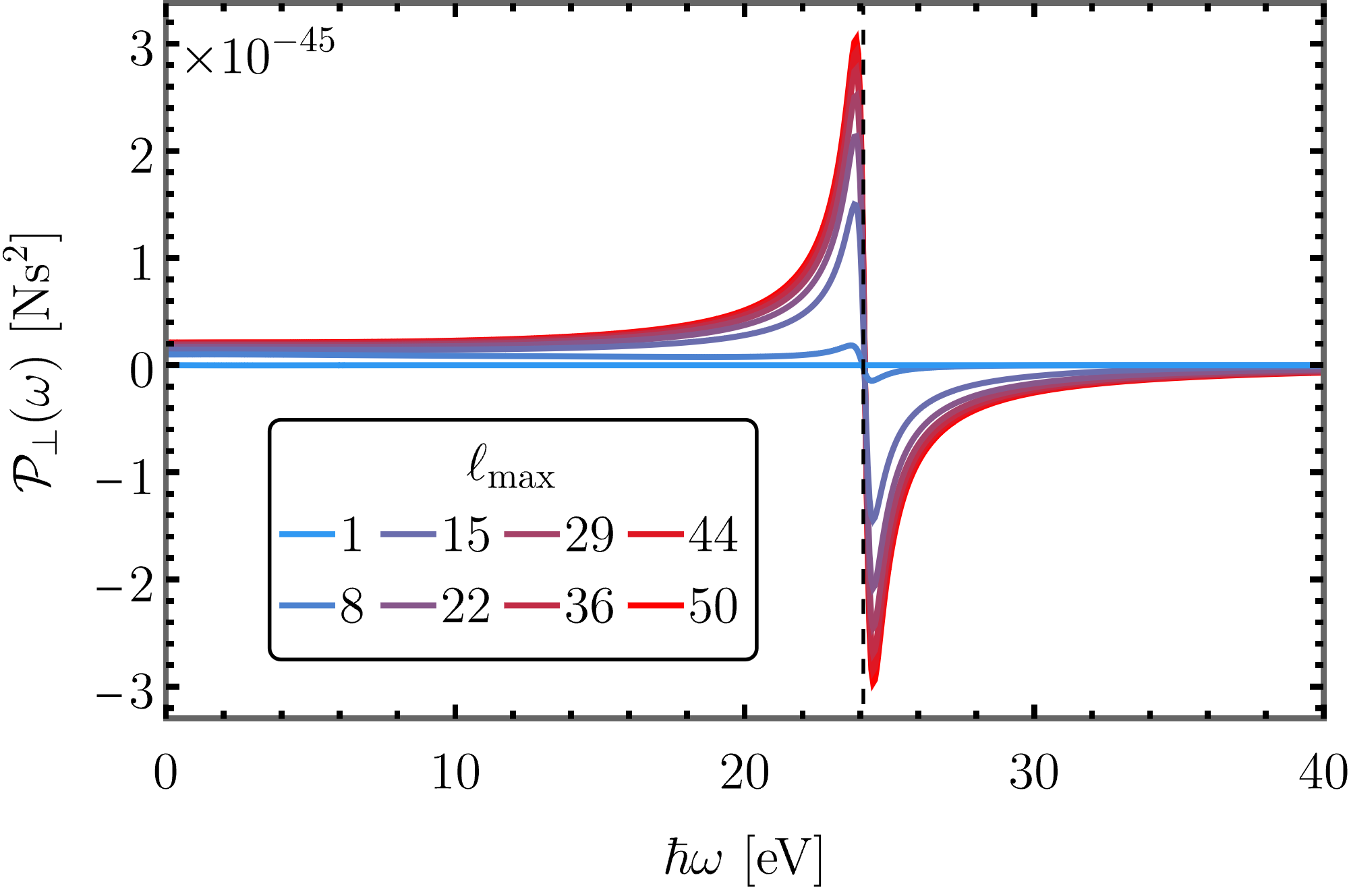}
    \caption{Same as Fig.~\ref{fig:Drude_convergence}, but for a single-Lorentz-oscillator test dielectric function
    \(\epsilon(\omega)=1+ A/(\omega_0^2-\omega^2-i\omega\Gamma)\),
    used to visualize the convergence around a single fixed material resonance.
    The resonance frequency is set to \(\hbar\omega_0=24.1~\mathrm{eV}\), chosen from the bismuth Lorentz parametrization in Table~\ref{tab:bismuth_params} (oscillator \(n=5\)).
    The vertical line marks \(\hbar\omega=\hbar\omega_0\).
    Convergence is also achieved for \(\ell_{\max}=47\).}
    \label{fig:Lorentzian_convergence}
\end{figure}

Figure~\ref{fig:Lorentzian_convergence} illustrates this convergence behavior and explicitly shows that the resonance consistently appears at $\omega_0$, as expected for a fixed Lorentz oscillator. As a guide to the eye, Fig.~\ref{fig:Lorentzian_convergence} includes a vertical reference line at $\hbar\omega=\hbar\omega_0=24.1~\mathrm{eV}$.

All results presented in the main text use $\ell_{\max}=50$. Convergence to within a relative error of $10^{-3}$ is already achieved for $\ell_{\max} \geq 47$, corresponding to approximately three significant digits of precision. In both test cases, $\ell_{\max}=50$ ensures that $\mathcal{P}_{\bot}(\omega)$ and the integrated momentum $\Delta P_{\bot}$ are fully converged.

\section{Longitudinal component of the Momentum Transfer}\label{app: momentum parallel}

For completeness, we briefly discuss the longitudinal (parallel to the trajectory of the electron) component of the linear momentum transfer, \(\Delta P_{\parallel}\), which corresponds to the momentum transferred along the propagation direction of the electron (\(\hat{\mathbf{z}}\)) defined in Fig.~\ref{fig:system}. As introduced in Section ~\ref{th}, the total linear momentum transferred to the nanoparticle can be decomposed into transverse and longitudinal components,
\(\Delta \mathbf{P}=\Delta P_{\bot}\,\hat{\mathbf{x}}+\Delta P_{\parallel}\,\hat{\mathbf{z}}\),
where \(\Delta P_{\bot}\) and \(\Delta P_{\parallel}\) are obtained by integrating the corresponding spectral densities \(\mathcal{P}_{\bot}(\omega)\) and \(\mathcal{P}_{\parallel}(\omega)\) over frequency.

Within the formalism developed in Section ~\ref{th A}, the longitudinal spectral density \(\mathcal{P}_{\parallel}(\omega)\) is evaluated using the same Maxwell stress tensor expression as the transverse component [Eq.~\eqref{eq: spectral density LMT}], with the unit vector \(\hat{\mathbf{n}}\) chosen along the \(\hat{\mathbf{z}}\) direction. Consequently, \(\mathcal{P}_{\parallel}(\omega)\) involves the same multipole expansions of the electromagnetic fields and the same truncation at \(\ell_{\max}\), and it obeys the same convergence criteria discussed in Appendix~\ref{app: multipole convergence}.

Although the present work focuses on the transverse momentum transfer because of its direct relevance to nanoparticle manipulation, the longitudinal component is of interest in its own right, as it is more closely connected to energy and momentum exchange along the electron trajectory and is therefore related to electron energy loss spectroscopy (EELS).

\begin{figure}[h!]
    \centering
    \includegraphics[width=0.9\linewidth]{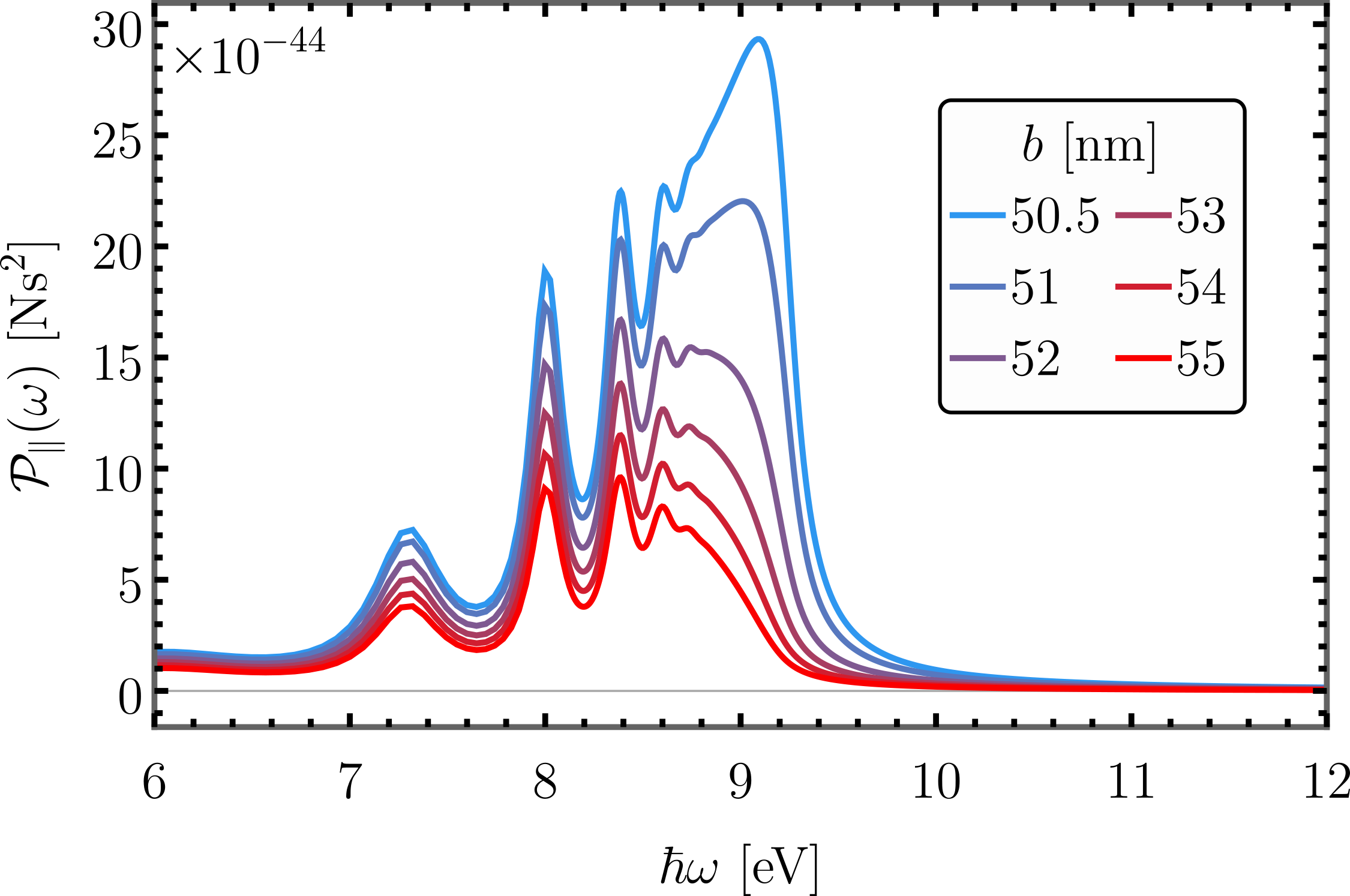}
    \caption{Representative example of the spectral density of the longitudinal linear momentum transfer \(\mathcal P_{\parallel}\) for a spherical Drude-like aluminum nanoparticle with $a=50$ nm, $b=50.5$ nm and $v=0.58c$. While \(\mathcal P_{\parallel}\) is more directly related to energy-loss processes probed by EELS, this figure demonstrates the ability of the present implementation to compute all components of the linear momentum transfer within a unified, fully converged formalism.}
    \label{fig:LongitudinalMomentum}
\end{figure}

Figure~\ref{fig:LongitudinalMomentum} shows a representative example of the spectral density of longitudinal momentum transfer \(\mathcal P_{\parallel}\) for a spherical nanoparticle, calculated for a Drude-like aluminum spherical NP of radius $50$~nm. The numerical implementation as in the transverse-momentum calculations is discussed in the main text. 

%%%%%%%%%%%%%%%%%%%%%%%%%%%%%%%%%%%%%%%%%%%%%%%%%%%%%%%%%%%%%%%%%%%%%%%%
\section{Dielectric function of bismuth}\label{app: dielectric Bi}
%%%%%%%%%%%%%%%%%%%%%%%%%%%%%%%%%%%%%%%%%%%%%%%%%%%%%%%%%%%%%%%%%%%%%%%%%

The dielectric response of bismuth is modeled using a sum of nine Lorentz oscillators, yielding
\begin{equation}
    \epsilon(\omega) = 1 + \sum_{n=1}^{9} 
    \frac{A_n}{\omega_{n}^2 - \omega^2 - i\omega\Gamma_n} \,,
\end{equation}
where $\omega_n$ and $\Gamma_n$ are the resonance frequency and damping coefficient of the $n$th oscillator, respectively, and $A_n$ is the corresponding oscillator strength. All energies are expressed in electronvolts (eV), with $\hbar$ absorbed for brevity.

The parameters used in this model are listed in Table~\ref{tab:bismuth_params}. This set reproduces the frequency-dependent dielectric response reported in Ref.~\cite{werner}.
\begin{table}[h!]%The best place to locate the table environment is directly after its first reference in text
\caption{\label{tab:bismuth_params}%
Lorentz-oscillator parameters used for the dielectric function of bismuth.
}
\begin{ruledtabular}
\begin{tabular}{cccc}
$n$ & $\hbar\omega_{n}$ (eV) & $\hbar\Gamma_{n}$ (eV) & $\hbar^2 A_{n}$ (eV$^{2}$) \\
\hline
1 & 0.0  & 0.087  & 108.4 \\
2 & 3.9  & 1.800  & 23.6  \\
3 & 6.6  & 4.500  & 50.2  \\
4 & 11.3 & 6.200  & 44.8  \\
5 & 24.1 & 0.600  & 6.6   \\
6 & 27.5 & 1.900  & 13.2  \\
7 & 40.2 & 9.600  & 99.3  \\
8 & 55.8 & 21.700 & 326.8 \\
9 & 84.2 & 48.400 & 373.1 \\
\end{tabular}
\end{ruledtabular}
\end{table}

%%%%%%%%%%%%%%%%%%%%%%%%%%%%%%%%%%%%%%%%%%%%%%%%%%%%%%%%%%%%%
\section{Numerical implementation}\label{app:numerical}
%%%%%%%%%%%%%%%%%%%%%%%%%%%%%%%%%%%%%%%%%%%%%%%%%%%%%%%%%%%%%%

The numerical implementation of the expressions developed in this work is openly available on GitHub~\cite{LMTRepo}. 
To reproduce Figs.~\ref{fig: Al_LMED} and~\ref{fig:DPDrudeAlBig}, the repository must be built locally, after which the simulations can be executed using
\begin{verbatim}
./DrudeAl_LMTsolver -a 50 --vscan
\end{verbatim}
or
\begin{verbatim}
./DrudeAl_LMTsolver -a 50 --bscan
\end{verbatim}

Each execution generates an output directory organized by material, nanoparticle radius, and execution timestamp. 
For a given run, the results are separated into folders corresponding to velocity scans at fixed impact parameter and impact parameter scans at fixed electron velocity. 
The velocity-scan results are further grouped according to the maximum multipole order, \(\ell_{\max}\), and include the computed transverse linear momentum transfer together with the associated numerical error estimates. 
Additional files record the multipolar convergence analysis. 
Each simulation directory also contains a text file summarizing the input parameters and metadata associated with the run.

Further details on compilation, runtime parameters available, and input options are provided in the repository \texttt{README} file.

%%%%%%%%%%%%%%%%%%%%%%%%%%%%%%%%%%%%%%%%%%%%%%%%%%%%%%
\section{Mie resonances versus EELS and CL}
\label{app: ext and scat mie}
%%%%%%%%%%%%%%%%%%%%%%%%%%%%%%%%%%%%%%%%%%%%%%%%%%%%%%

\begin{figure}[h!]
    \centering
    \includegraphics[width=0.9\linewidth]{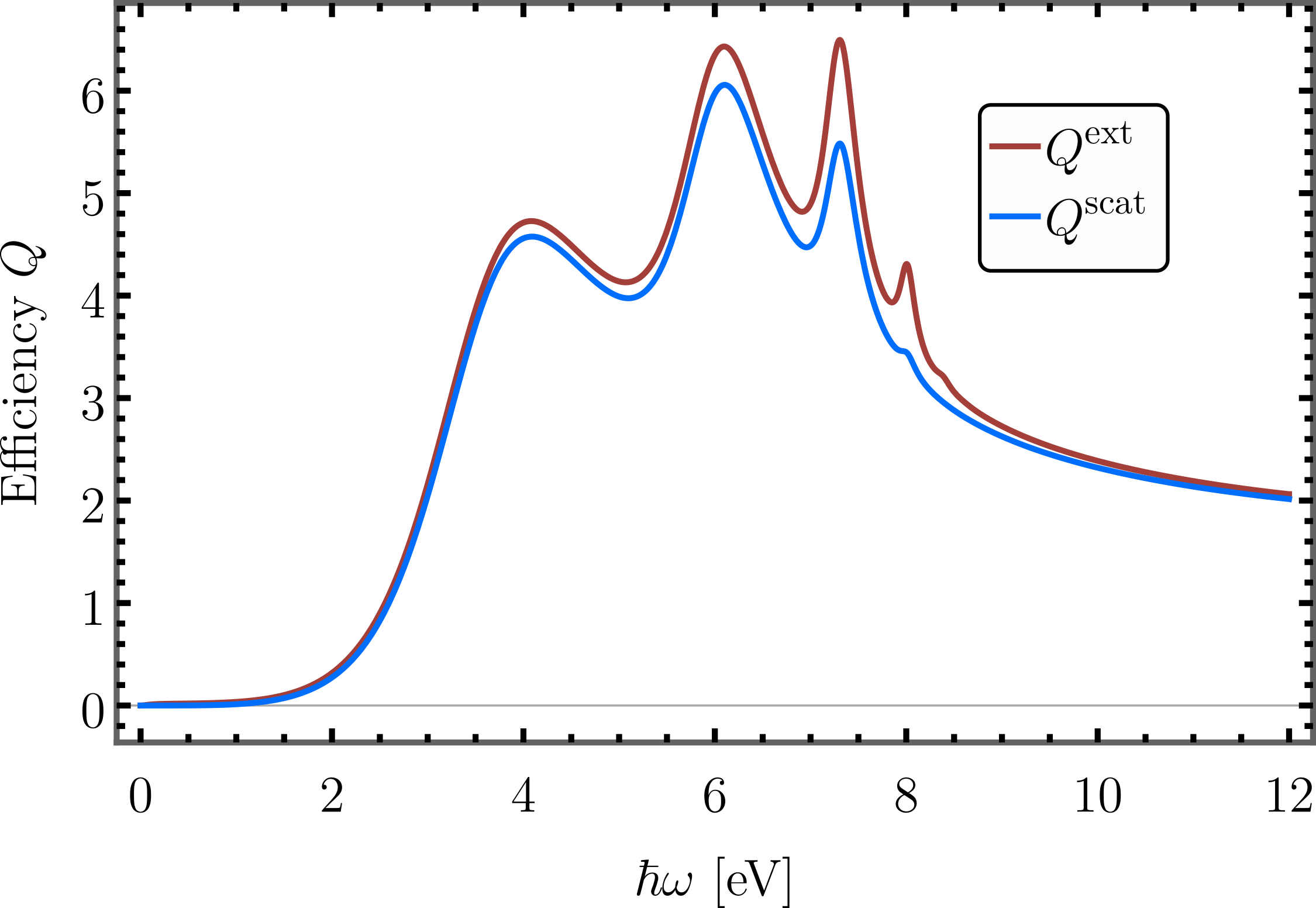}
    \caption{Mie scattering and extinction efficiencies for a Drude-like aluminum nanoparticle with radius $a=50$~nm. The amplitudes of the dominant resonances do not directly correspond to the peaks observed in the transverse spectral momentum-transfer densities, which display additional multipolar contributions and clearly reveal the surface plasmon resonance at $\hbar\omega_s$.}
    \label{fig:Mie_coefficients}
\end{figure}

Figure~\ref{fig:Mie_coefficients} presents Mie scattering and extinction efficiencies for a Drude-like aluminum nanoparticle with radius $a=50$~nm. Although these quantities are commonly used to identify dominant optical resonances, we find that the amplitudes of the strongest resonances in the Mie efficiencies do not directly correspond to the peaks observed in the transverse spectral density of linear momentum transfer shown in Figs.~\ref{fig: Al_LMED}(c) and~\ref{fig: Al_LMED}(d).

The momentum-transfer spectral density exhibits a richer spectral structure, including additional resonances that are absent or weak in the Mie efficiencies. This discrepancy arises because the momentum-transfer spectrum depends not only on the intrinsic optical response of the nanoparticle, encoded in the Mie coefficients, but also on the spatial overlap between the electromagnetic field generated by the swift electron and the multipolar modes of the nanoparticle. 

As a result, multipolar excitations that weakly contribute to far-field scattering may be efficiently driven by the near field of the electron and become prominent in the momentum-transfer spectrum. In particular, the spectral density clearly reveals the surface plasmon resonance at $\omega_s$, corresponding to the asymptotic mode $\ell \rightarrow \infty$ of the nanoparticle, a feature that is comparatively less pronounced in far-field Mie efficiencies.

\begin{figure}[h!]
    \centering
    \includegraphics[width=0.9\linewidth]{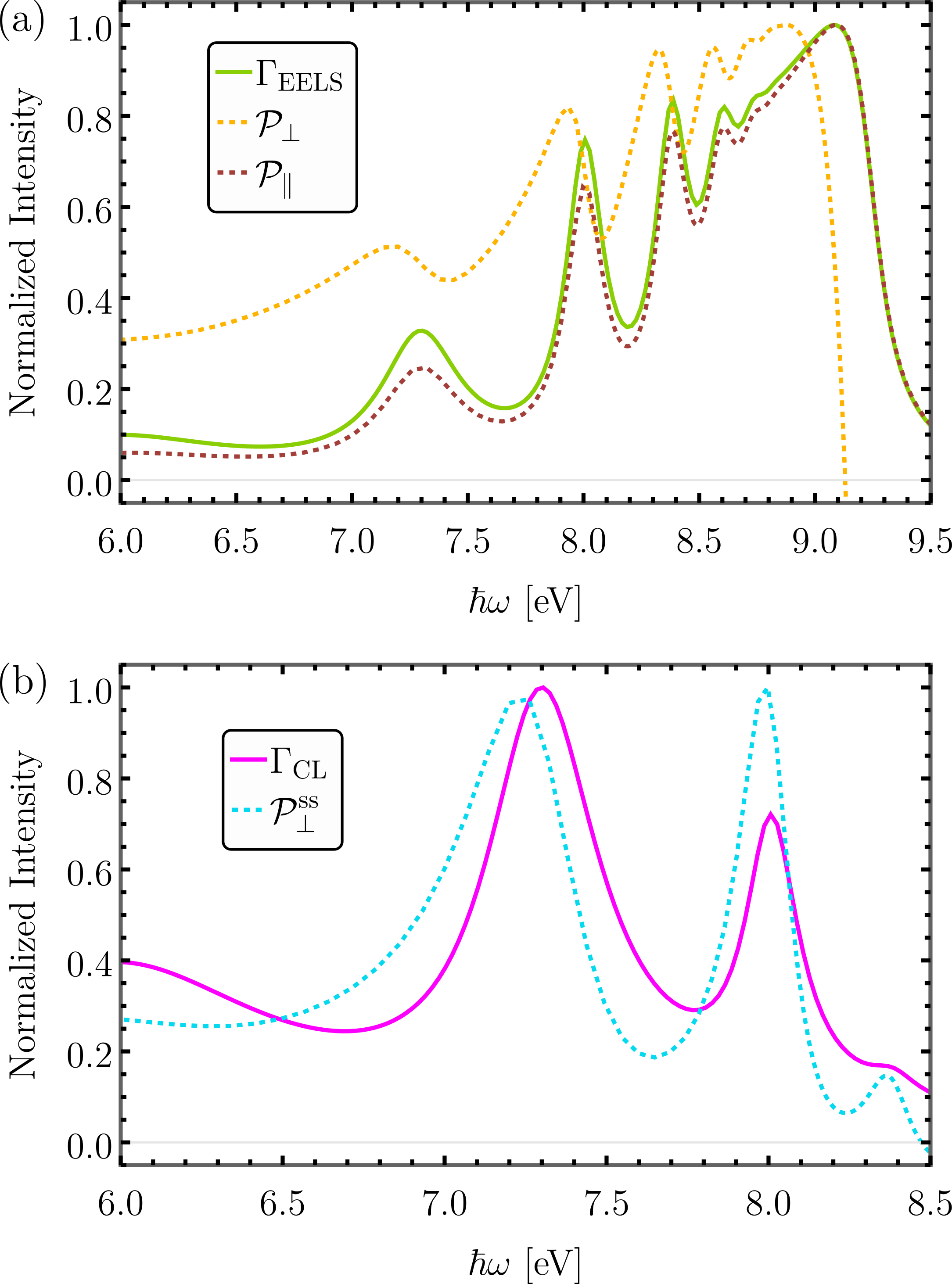}
    \caption{Comparison between momentum-transfer spectral density and electron-beam observables for a Al NP with $a=50$~nm, $b=50.5$~nm, and $v=0.58c$. 
    (a) $\mathcal{P}_{\parallel}(\omega)$, $\mathcal{P}_{\bot}(\omega)$, and the EELS probability $\Gamma_{\mathrm{EELS}}(\omega)$ (all normalized to their respective maxima). The main resonances occur in the same frequency regions. 
    (b) Scattered--scattered $\mathcal{P}_{\bot}^{\mathrm{ss}}(\omega)$ contribution together with the CL probability $\Gamma_{\mathrm{CL}}(\omega)$. The peaks of $\mathcal{P}_{\bot}^{\mathrm{ss}}(\omega)$ qualitatively align with those of $\Gamma_{\mathrm{CL}}(\omega)$.}
    \label{fig:EELS_CL}
\end{figure}

For comparison, Fig.~\ref{fig:EELS_CL}(a) shows $\mathcal{P}_{\parallel}(\omega)$, $\mathcal{P}_{\bot}(\omega)$, and the EELS probability $\Gamma_{\mathrm{EELS}}(\omega)$ for the same parameters. The dominant resonances appear in similar frequency regions, indicating that EELS captures spectral features closely related to the longitudinal and transverse momentum-transfer spectra.

Figure~\ref{fig:EELS_CL}(b) compares the scattered--scattered contribution, $\mathcal{P}_{\bot}^{\mathrm{ss}}(\omega)$, with the CL probability $\Gamma_{\mathrm{CL}}(\omega)$. The peaks of $\mathcal{P}_{\bot}^{\mathrm{ss}}(\omega)$ occur in the same frequency regions as those of $\Gamma_{\mathrm{CL}}(\omega)$, consistent with the radiative character of both quantities.

These comparisons further clarify the relation between momentum transfer and conventional electron-beam observables, and complete the analysis presented in this Appendix.

\vspace{0.1cm}

\newpage
%
%
% REFERENCES
%
%

\bibliography{references}

\end{document}